\documentclass[preprint,12pt]{elsarticle}

\usepackage{amsmath}
\usepackage{amsfonts}
\usepackage{amssymb}
\usepackage{graphicx}
\usepackage{geometry}
\usepackage{setspace}
\usepackage[latin1]{inputenc}
\usepackage{fancyhdr}
\newcommand{\dd}{\mathrm{d}}
\newcommand{\vv}[1]{\overrightarrow{#1}}
\usepackage{color}

\journal{International Journal of Heat and Mass Transfer}

\begin{document}
\begin{frontmatter}
 
\title{\textbf{Thermal chaotic mixing: comparison of constant wall temperature and constant heat flux boundary conditions}}

\author[ky]{Kamal El Omari\corref{cor1}}
\ead{kamal.elomari@univ-pau.fr}
\author[ky]{Yves Le Guer}
\ead{yves.leguer@univ-pau.fr}

\address[ky] {Laboratoire des Sciences de l'Ingénieur Appliquées à la Mécanique et au Génie Electrique (SIAME)\
	{Université de Pau et des Pays de l'Adour (UPPA)}	\\
	{IFR, rue Jules Ferry, BP. 7511, 64075 Pau cedex - France}}

\cortext[cor1]{Corresponding author, Tel: +33 559 407 147, Fax: +33 559 407 160}

\begin{abstract}
In a recent paper (El Omari and Le Guer, IJHMT, 53, 2010) we have investigated mixing and heat transfer enhancement in a mixer composed of two circular rods maintained vertically in a cylindrical tank. The rods and tank can rotate around their revolution axes while their surfaces were maintained at a constant temperature. In the present study we investigate the differences in the thermal mixing process arising from the utilization of a constant heat flux as a boundary condition.  The study concerns a highly viscous fluid with a high Prandtl number $Pr = 10,000$ for which this chaotic mixer is suitable. Chaotic flows are obtained by imposing temporal modulations of the rotational velocities of the walls. By solving numerically the flow and energy equations, we studied the effects of different stirring protocols and flow configurations on the efficiency of mixing and heat transfer. For this purpose, we used different statistical indicators as tools to characterize the evolution of the fluid temperature and its homogenization. Fundamental differences have been reported  between these two modes of heating or cooling: while the mixing with an imposed temperature results in a homogeneous temperature field, with a fixed heat flux we observe a constant difference between the maximal and minimal temperatures that establish in the fluid; the extent of this difference is governed by the efficiency of the mixing protocol.
\end{abstract}


\end{frontmatter}

%\linenumbers

\textbf{Nomenclature}
\begin{tabbing}
 \hspace{1.5cm}\=\kill
$C_p$ \>  heat capacity ($J/kg.K$)\\ 
$k$ \> thermal conductivity ($W/m.K$)\\ 
$L$ \> wall characteristic length ($m$)  \\ 
 $p$\> pressure ($Pa$)\\ 
$\dot q$ \> surface heat flux ($W/m^{2}$) \\ 
$R_3$ \>  tank radius ($m$) \\
$R_1, R_2$ \> rod radii ($m$)  \\ 
$t$ \>  time ($s$)\\ 
$T$ \> temperature ($K$) \\ 
$U$ \>  tangential velocity ($m/s$)\\ 
$\vec U$ \>  velocity field ($m/s$)\\ 
%%%%%%%%%%%%%%%%%%%%%%%%%%%%%%%%%%%%%%%%%%%%%%
\textbf{Dimensionless numbers} \\ 
$Nu$ \>  Nusselt number\\ 
$Pe$ \> Péclet number \\ 
$Pr$ \>  Prandtl number\\ 
$Re$\>  Reynolds number\\ 
%$T^*$ \>  dimensionless temperature\\ 
$X$ \>  rescaled dimensionless temperature\\
%%%%%%%%%%%%%%%%%%%%%%%%%%%%%%%%%%%%%%%%%%%%
\textbf{Greek symbols} \\ 
$\varepsilon$\>  eccentricity ($m$)\\ 
$\rho$ \>  fluid density ($kg/m^{3}$)\\ 
$\sigma$ \>  standard deviation\\ 
$\tau$ \> period of modulation ($s$) \\ 
$\overset{=}{\tau}$ \>  viscous stress tensor ($Pa$)\\ 
$\Omega$ \> angular velocity ($rad/s$) \\ 
%%%%%%%%%%%%%%%%%%%%%%%%%%%%%%%%%%%%%%%%%%%%%%%
\textbf{Subscript} \\ 
$c$ \>  cell\\ 
 $m$\>  mean\\ 
$f$\>  face of a cell\\ 
\end{tabbing}

%%%%%%%%%%%%%%%%%%%%%%%%%%%%%%%%%%%%%%%%%%%%%%%%%
\section{Introduction}

Mixing processes are widely encountered in many practical engineering domains where enhancements of the heat, mass and momentum transfer are required. In the present study, mixing was achieved through the existence of a laminar chaotic flow, which ensures the efficient stretching and folding of the material lines (here for a two-dimensional (2D) flow). The need for chaotic mixing is particularly interesting when high viscosity and/or shear-sensitive fluids are concerned \citep{ottino1989}. In this case, classical laminar 2D time-independent flows are unable to give a good mixing performance and as a consequence, heat transport from the wall will be ineffective \citep{aref2002}. Such a situation exists for molten polymers or polymer blends \citep{jana2004}, in food processing \citep{metcalfe2009}  or are encountered for the formation of highly concentrated emulsions \citep{caubet2010}. Also, the processing of highly viscous polymer molecule networks or biological fluids in classical turbulent flow mixers can cause the degradation of these fluids in the high-shear regions of the flow. Besides, undesired or ill-defined fluid structures may be obtained.

Chaotic flows were successfully used for heat transfer \citep{chang1994}, especially in twisted-pipe heat exchangers \citep{acharya2001}. Even if the flow in these pipes is steady, the chaotic character is obtained by elbows succession, where at sufficiently high Dean numbers, secondary vortices appear and promote chaotic advection \citep{peerhossaini1993,mokrani1997}. This type of mixer is efficient at intermediate Reynolds numbers, roughly for $40<Re<2000$, and is thus not suitable for highly viscous fluids such as the one addressed in the present study.

The thermal chaotic mixing has some particularities compared to the mixing of concentration (dye for example). Even if concentration and temperature are both passive scalars, their mixing require different strategies. The origin of this difference is the location of the scalar source.
In the case of mixing of concentration, the two constituents are present in the mixer at the start of the process. Whereas, for the thermal mixing, the fluid is initially at a uniform temperature (a cold temperature for example in the case of a heating process) and the heat source is located at the mixer boundaries (in the absence of volumetric heating as the Ohm's effect). 
Two different situations are then encountered depending on the type of the boundary condition imposed at the mixer walls: (\textit{i}) if a fixed temperature is imposed, the amount of energy extracted from these walls by the fluid (parietal heat flux rate) depends on the ability of the flow to generate fluid trajectories perpendicularly to them and to mix this hot fluid with the colder one far from the boundaries, (\textit{ii}) if a fixed flux is imposed at the walls, the rate of heat flux is then known and the behavior of the flow will impact the repartition of this heat between the different zones of the fluid. However, in both situations, the flow in the mixer must insure a good transport from the walls towards the center.

In the case of concentration mixing no source is present since the second constituent is initially injected in the mixer as a small blob in the case of dye mixing for example, or it occupies a larger portion of the mixer space in the case of the blend of two fluids. As a consequence, the mixing strategies classically developed for concentration mixing may not be successful for thermal mixing if it neglect the mixing from and near the walls.

When mixing a blob of dye in an enclosed fluid, \citet{gouillart2008} showed that the presence of the no-slip wall of a non-rotating tank slows down the rate of mixing, while the rate is exponential when this tank is rotating \citep{gouillart2010}. This is due to the presence of parabolic points in the non-rotating wall which brings unmixed fluid to their zone of interest, i.e. the center of their mixer. When the tank is rotating, the central zone is isolated from the peripheral unmixed zone and the rate of the scalar decay is exponential as expected for a fully chaotic flow. When dealing with thermal mixing as it is the case in the present work, the situation is reversed since the scalar (temperature) needs to be driven from the wall. We showed in \citep{elomari2010a} that the presence of parabolic points on the boundaries is a necessary condition in order to mix fluid from the walls and bring it towards the mixer center.

The mixer geometry used here was derived from the classical eccentric cylinder geometry (Journal bearing flow) considered by different authors \citep{ghosh1992,ganesan1997,lefevre2003,mota2007}. This geometry produces a large recirculation bubble, which is difficult to avoided even with a careful choice of stirring protocols. In contrast, the two-rod mixer studied here is suitable for obtaining a full chaotic flow without KAM regions. It is  similar to the geometry of the two-roll-mills studied in the literature by  \citet{price2003} or \citet{young2007} and \citet{chiu2009}. This geometry was studied in details by \citet{jana94} under the name of ``Vortex mixing flow''. They used numerical and experimental tools to study the mixing of a scalar (dye in the experiments) by rotating alternately the two rods in a fixed tank (the outer cylinder). In our case, the tank has the ability to rotate. We previously studied this mixer in the case of heating of a Newtonian fluid using a fixed temperature at the mixer walls \citep{elomari2009a,elomari2010a}. We extended these studies to the case of shear-thinning and shear-thickening fluids \citep{elomari2010b}.

The objective of the present work is to analyze the combined effects of mixing and heat transfer for the particular wall boundary condition of constant heat flux (i.e. the Neumann condition). In the authors' previous studies \citep{elomari2009a,elomari2010a,elomari2010b}, mixing and heat transfer were studied for the constant wall temperature boundary condition (i.e. the Dirichlet condition) in the same mixer. In such a case, the imposed wall temperature represents an asymptotic limit for the evolution of the mean fluid temperature, and the rapidity to reach this limit is controlled by the efficiency of the mixing. In contrast, for constant wall heat flux, there is no asymptotic limit for the evolution of the mean fluid temperature because its evolution is prescribed by the imposed parietal heat flux density.  As a consequence, the stirring strategy and the choice of mixing indicators must be selected in agreement with the type of wall heating considered.  The differences in mixing induced by the choice of boundary condition (constant wall heat flux versus constant wall temperature) will be highlighted throughout this work. 

\section{Problem formulation}
\label{problemformulation}

\subsection{Geometry of the two-rod rotating mixer}
\begin{figure}[htbp]
 \centering
 \begin{center}
 \includegraphics[width=0.5\textwidth]{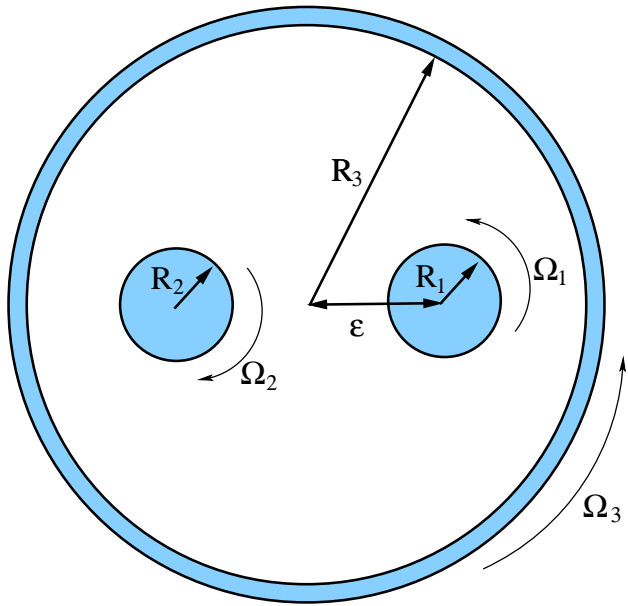}
\end{center}
 \caption{Sketch of the two-rod mixer.}
 \label{fig:mixer}
\end{figure}

A sketch of the mixer studied is presented in Fig. \ref{fig:mixer}. It is composed of two circular rods of equal radii, which are maintained vertically inside a cylindrical tank (a bounded domain). The tank and the rods are heated and can rotate around their respective revolution axes. The rods and the tank rotate alternately or with a continuous modulation of velocities. They can also have different directions of rotation. We have shown for a Newtonian fluid \citep{elomari2009a,elomari2010a} that this two-rod mixer is suitable for obtaining full chaotic flow without KAM regions, which is particularly interesting for industrial applications.
The geometry of the flow domain is characterized by the radii of the rod and the cylindrical tank, which are respectively: $R_1 = R_2 = 10\ mm$ and $R_3 = 50\ mm$. The eccentricity of the rods is set to the value of $\varepsilon= 25\ mm$. 
The choice of $\varepsilon= R_3/2$, i.e. the rods located at the midpoint between the center of the tank and its boundary, was found to be close to the optimum in our previous study \citep{elomari2010a}. This value is also near to the optimum of eccentricity found by \citet{jana94} for a radii ratio  $R_3 /R_1=4$ and for a fixed tank and alternately rotating rods.

\subsection{Fluid properties}
\label{fluidproperties}
The Newtonian fluid considered in this study has the thermophysical properties listed in Table~\ref{tab:fluid}.

\begin{table}[tbh]
 \begin{center} 
\begin{tabular}{l l}\hline
Dynamic viscosity ($\mu$)& 1.5 $Pa.s$\\
Density ($\rho$) & 990 $kg.m^{-3}$\\
Thermal conductivity ($\lambda$) & 0.15 $W. m^{-1} K^{-1}$\\
Specific heat ($C_p$) & 1000 $J. kg^{-1} K^{-1}$\\
Prandtl number ($Pr$) & $10^4$\\
P\'eclet number ($Pe$) & $16,584$\\
\hline
 \end{tabular}  
\caption{Fluid properties.\label{tab:fluid}}
\end{center}
\end{table}

Considering these properties, the thermal diffusivity $\alpha$ is equal to $1.515\times 10^{-7} \ m^2.s^{-1}$. The maximum angular wall velocities are fixed to: $\Omega_1 = \Omega_2 = 30$ rpm for the rods and $\Omega_3 = 6$ rpm for the outer tank. The tangential wall velocity is then the same, and equal to $U = 31.41\ mm.s^{-1}$. A characteristic Reynolds number for this flow can be evaluated as:

\begin{equation}
 Re = \dfrac{\rho\ U\cdot 2\cdot(R_3 - R_1)}{\mu} = 1.66
 \label{eq:re}
\end{equation}

In our study, the Péclet number Pe is large (see Tab.~\ref{tab:fluid}) hence the need to accelerate the mixing of the temperature by advection is evident.

\subsection{Flow parameters: stirring protocols}
\label{strirringprot}
Chaotic mixing flows are produced by varying the angular velocity of the rods and tank in time using a sine-squared waveform. 
The stirring protocols studied are defined by two parameters: the respective direction of the rotation between the rods and tank, and the size of the period of the velocity modulation. Other parameters, such as the ratio between the maximum velocities of the rods and tank, and the amplitude of the angular velocity modulation, are not considered here. The three possible flow configurations giving rise to different flow topologies are specified in Table \ref{tab:stirringconf}. The rods are always rotating in phase. The symbol ($+$) indicates a counter--clockwise direction and ($-$), a clockwise direction.

\begin{table}[tbh]
 \begin{center}
\begin{tabular}{c c c c}\hline
Flow configuration & Rod 1 & Rod 2 & Tank \\\hline
A        &  ($+$)      &  ($-$)      &  ($+$) \\
B        &  ($+$)      &  ($+$)      &  ($+$) \\
C        &  ($-$)      &  ($-$)      &  ($+$) \\\hline
\end{tabular}  
\caption{The three possible flow configurations related to the sign of angular velocity. \label{tab:stirringconf}}
\end{center}
\end{table}   

\begin{figure}[tb]
 \centering
 \begin{center}
 \includegraphics[width=0.7\textwidth]{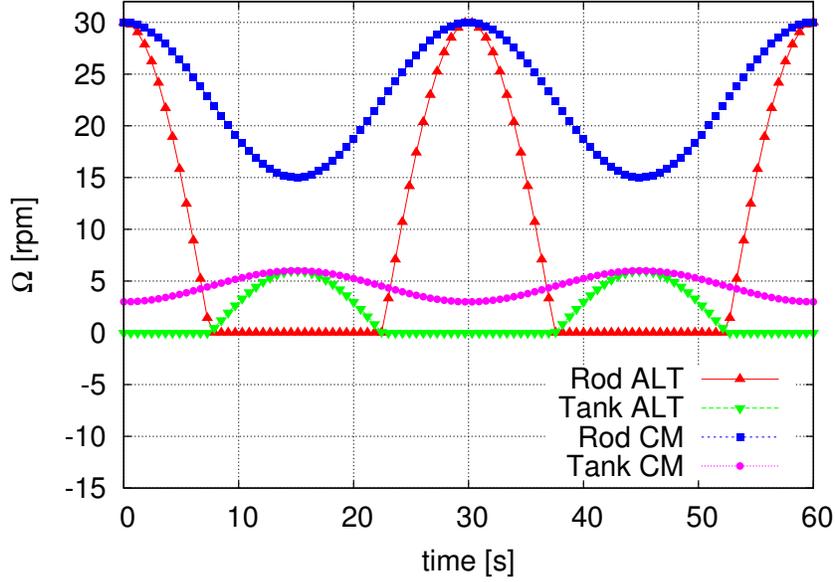}
\end{center}
 \caption{The temporal modulation of the angular velocities of the rods and tank for the continuous (CM) and non-continuous (ALT) stirring cases. Here the modulation period, $\tau$, is $30\ s$.}
 \label{fig:modulation}
\end{figure}

In order to facilitate comparisons, the same types of temporal modulation are considered as in our previous work \citep{elomari2010a}. These are, for chaotic flows: a continuous modulation denoted CM (a sine-squared variation of the wall velocities) and a non-continuous modulation (alternated) denoted ALT, i.e. the rods are stopped together for half a period while the outer tank is rotating; for the next half-period the contrary applies. Three periods of modulation  $\tau$ are considered here : $15\ s$, $30\ s$ and $60\ s$.
These stirring protocols are illustrated in Fig. \ref{fig:modulation}. The maximum angular velocity is the same for the two types of modulation. The rods and the tank follow sine-squared modulations, they are detailed in \citep{elomari2010a}.  In our modeling, we take into account inertial effects as the flow considered for this tangential wall velocity does not satisfy the quasi-steady hypothesis.

%As stated above, the maximum angular wall velocities are fixed to: $\Omega_1 = \Omega_2 = 30$ rpm for the rods and $\Omega_3 = 6$ rpm for the outer tank in such a way as to have the same  tangential wall velocity equal to $U = 31.41\ mm.s^{-1}$.

\subsection{The constant wall heat flux boundary condition}
\label{wallboundary}
The Neumann boundary condition (constant wall heat flux) is imposed on the rods and tank walls. As mentioned above, in previous studies \citep{elomari2009a,elomari2010a}, we considered the constant wall temperature boundary condition (i.e. Dirichlet condition). This new boundary condition (Neumann condition) imposes to reconsider the objective of the mixing and heat transfer optimization problem (i.e. the optimal choice of the stirring protocols). 
Indeed, while in the case of a constant wall temperature the final mean temperature is asymptotically bounded by the parietal temperature, in the case of a constant wall heat flux the mean temperature $T_m$ will instead increase linearly with time and be directly proportional to the parietal flux. Effectively, for a 2D mixer we have a slope of:
\begin{equation}
 \dfrac{d T_m}{d t} = \dot{q} \left( \dfrac{S}{\rho\;\ C_p\ V}\right) 
 \label{eq:qpoint}
\end{equation}
where V is the volume of the fluid and S the outer surface of the mixer. 

Thus, the statistical values to follow in order to characterize the efficiency of the process are not the same ones used previously. Instead of following the mean temperature, we follow the temporal evolutions of the difference between the maximum and minimum temperatures or the variation of one of them compared to the mean temperature (see section \ref{results}). 
Therefore, in the case of poor mixing, extremum temperatures ($T_{max}$ and $T_{min}$) should have high values in comparison with the case of a good mixing process. Higher temperatures will be encountered in the vicinity of the heating walls and may cause undesirable damages to the fluid.

\section{Computational modeling}

\subsection{The governing equations}\label{goveqnummeth}
The unsteady Navier-Stokes equations governing the flow of incompressible fluids and the continuity equation are considered in their integral form:
\begin{equation}
 \dfrac{\partial}{\partial t} \int_{V}\rho\ \vec U\ \dd V
   + \int_{S} \rho\ \vec U \vec U \cdot \vec n \ \dd S =
   \int_{V} -\vec\nabla p \ \dd V  
+ \int_{S} \overset{=}{\tau}\cdot \vec n \ \dd S  \label{eq:mvt}
\end{equation}
\begin{equation}
 \int_{S} \vec U \cdot \vec n \ \dd S =0
\label{eq:4}
\end{equation} 
where $\overset{=}{\tau}$, the viscous stress tensor for a Newtonian fluid:
\begin{equation}
    \overset{=}{\tau}= \mu \left( \overset{=}{\nabla}U+(\overset{=}{\nabla}U)^T \right)
\end{equation}

The integration is over a volume $V$ surrounded by a surface $S$ oriented by an outward unit normal vector $\vec n$. The equation of the energy conservation is considered in terms of the temperature:
\begin{eqnarray}
\dfrac{\partial}{\partial t} \int_{V}\rho\ C_p T\ \dd V
   + \int_{S} \rho\ C_p T\ \vec U \cdot \vec n \ \dd S &=&
   \int_{S} k\ \vec\nabla T\cdot \vec n \ \dd S
\label{eq:energy}
\end{eqnarray}
where $k$ is the thermal conductivity.
The temperature is non-dimensionalized by:
\begin{equation}
T^* = \dfrac{T-T_{ini}}{\Delta T}\qquad \text{with}\qquad \Delta T = \dfrac{\dot q_{ref}\  L}{k}
\end{equation}
$\Delta T$ is fixed to 1, thus, $\dot q_{ref} = \dfrac{k}{L}$ with $L = 2\:(R_3 -R_1)$.

At the beginning of the mixing process, the fluid is at a uniform initial temperature $T_{ini}$ and the heat flux at the different walls is fixed to $\dot q = 10\ \dot q_{ref}$.

\subsection{The numerical method}

The conservation equations (\ref{eq:mvt}, \ref{eq:4} and  \ref{eq:energy}) are solved by the means of an in-house code called Tamaris. This code has a three-dimensional unstructured finite volume framework that is applied to hybrid meshes. It is a direct extension to 3D of the 2D code used in our previous study \cite{elomari2010a}.   The cell shapes can be of different forms (tetrahedral, hexahedral, prismatic or pyramidal). The variable values ($\vec U$, p and T) are stored at cell centers in a collocated arrangement. This three-dimensional (3D) code can also deal with 2D computations (e.g. in ($\vec x,\vec y$) plan) without any change, by considering a single layer of computational cells (in $\vec z$ direction) and by neutralizing the top and bottom faces (with respect to $\vec z$). In the scope of this work, all of the computational meshes were generated using the open-source software Gmsh \citep{geuzaine2009}.
More details about the numerical method and the discetization procedures used are reported in the appendix A.
The code validation was undertaken in \cite{elomari2010b} and extended here in the appendix B with a supplementary experimental test case.

\section{Mixing indicators}
\label{mixingindicators}

In order to quantify the efficiency of the heating process for all the fluid inside the mixer, we used the following instantaneous measure: the standard deviation $\sigma$ of the dimensionless  temperature $T^*$. This quantity is defined as (the summation is made over all the mesh cells $c$ of area $A_{c}$):
\begin{equation}
 \sigma = \left[\dfrac{1}{\sum_{c} A_{c}}\sum_{c}\left( A_{c}(T^*_{c} -T^*_m)^2\right) \right]^{\frac{1}{2}}
\label{eq:monitor2}
\end{equation}
where $T^*_m$ in the non-dimensional mean temperature calculated as:
\begin{equation}
  T^*_m = \dfrac{1}{\sum_{c} A_{c}}\left(\sum_{c} A_{c} T^*_{c}\right) 
\label{eq:monitor1}
\end{equation} 

The standard deviation is an indicator of the level of homogenization of the scalar temperature inside the 2D tank \citep{elomari2010a}. The evolution of the mean temperature gives the total energy supplied to the fluid during the mixing process, which is directly proportional to the value of the wall heat flux when we consider the Neumann boundary condition (see section \ref{wallboundary}).

Another mixing indicator is the temperature scalar dissipation indicator, which is defined as:
\begin{equation}
\chi_g = \dfrac{1}{S_{tot}}\int_{S_{tot}} \lVert\vec\nabla T^*\rVert^2 \; \mathrm dS 
= \dfrac{1}{\sum_{c} A_{c}}\left(\sum_{c} A_{c}\lVert\vec\nabla T^*\rVert^2_c \right)
 \label{eq:chi}
\end{equation}

where $S_{tot}$ is the total surface of the fluid in the mixer. This indicator illustrates the mechanism of scalar gradient production and destruction in the fluid.

Heat exchange at a wall (rod or tank) is characterized here by the mean Nusselt number (we recall that $\dot q = 10\ \dot q_{ref}$, see section \ref{goveqnummeth}):
\begin{equation}
\overline{Nu} = \dfrac{1}{S_{wall}}\int_{S_{wall}} \dfrac{\dot q \ L}{k (T^*_{wall}-T^*_m)}  \mathrm dS  =  \dfrac{1}{S_{wall}}\int_{S_{wall}} \dfrac{10}{T^*_{wall}-T^*_m}  \mathrm dS
\end{equation}

\section{Results and discussion}
\label{results}
The mesh of the two-rod mixer used in this work  contains $10680$ computational cells and uses regular quadrilateral cells near the walls to enhance the resolution of the boundary layers (the figure of this mesh was shown in \citep{elomari2010b} so it is not reproduced here). This mesh was adopted after a grid size-dependence study as detailed in \citep{elomari2010b} which was extended to the case of prescribed heat flux.

In this section we describe the obtained temperature field for the three stirring protocols studied and we show the impact of these protocols on the defined mixing indicators, temperature extrema and distributions (PDF). A particular attention is paid to the differences with the case of constant temperature boundary condition studied in \cite{elomari2010a}.

% \begin{figure}[htbp]
%  \centering
%  \begin{center}
%  \includegraphics[width=0.5\textwidth]{./figs/mesh.eps}
% \end{center}
%  \caption{Computational mesh of the two-rod mixer.}
%  \label{fig:mixer_mesh}
% \end{figure}

\subsection{Temperature patterns}

\begin{figure}[tbh]
 \centering
 \begin{center}
 \includegraphics[width=0.99\textwidth]{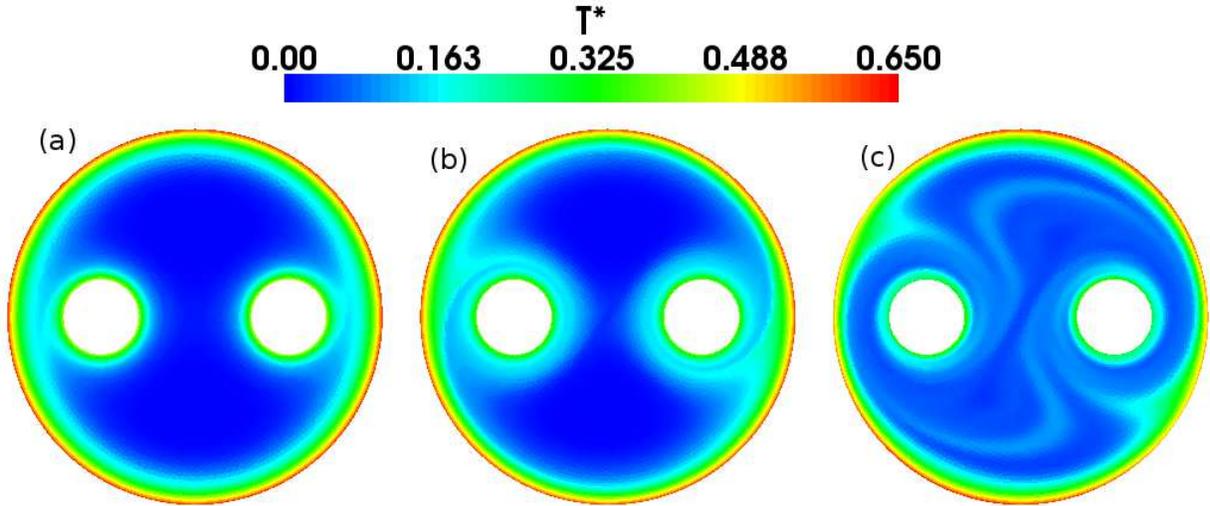}
\end{center}
 \caption{Dimensionless temperature fields at time $t = 120\ s$ for the flow configuration C and for: (a) non-modulated (NM), (b) continuously modulated (CM) and (c) alternated (ALT) stirring protocols (period of modulation $\tau=30\ s$).}
 \label{fig:snapshot_temp}
\end{figure}

In Fig.~\ref{fig:snapshot_temp}, the dimensionless temperature fields are shown for the stirring condition C of Table \ref{tab:stirringconf} and for three cases of the modulation of the wall angular velocity. After four periods ($t= 120\ s$), we find that the temperature fields of the non-modulated (NM) and continuously  modulated (CM) cases are quite similar. They follow the shape of the roughly steady-state streamline patterns and present large zones of unmixed cold fluid. Due to the continuous rotation of the walls in these two cases, there are permanent streamlines parallel to the boundaries that inhibit heat transport by advection to the center of the mixer. The flow patterns in the two-rod mixer were exhaustively presented and discussed in our previous paper \citep{elomari2010a} dealing with imposed temperature as a boundary condition, so they are not reproduced here. Unlike the NM and CM cases, the alternated modulation (ALT) case allows hot fluid strips to penetrate towards the center of the mixer. Indeed, when one of the walls is at rest, stagnation points form at its boundary and give rise to such a hot fluid streams. 
As said earlier, when the walls are non-rotating, parabolic points appear in them, and by alternated rotation they take place successively in the tank and rods surfaces. In Figure \ref{fig:snapshot_temp}(c), we can indentify two hot streams originating from the tank boundary, carried by the rotation of the rods (at $t= 120\ s$, the rods are rotating at their maximum velocity while the tank is at rest, see Fig. \ref{fig:modulation}). Also, with an alternated rotation, quite different flow topologies with crossing streamlines succeed each other in the mixer, and this feature is the key of efficient chaotic mixing \citep{ottino1989}.

When comparing the temperature fields obtained for the fixed temperature boundary condition to those of Fig. \ref{fig:snapshot_temp}, we find exactly the same patterns, but with different scales. For the former case, the maximum value is 1 and is always present in the field, while the minimum value evolves continuously towards 1. For the fixed flux boundary condition, both extremums are evolving. These observations and the impact of the mixing efficiency on them are detailed in section \ref{sec:extremums}.

\subsection{Influence of the wall boundary condition on the temperature homogenization}

\begin{figure}[tbh]
 
 \begin{center}
  \includegraphics[width=0.49\textwidth]{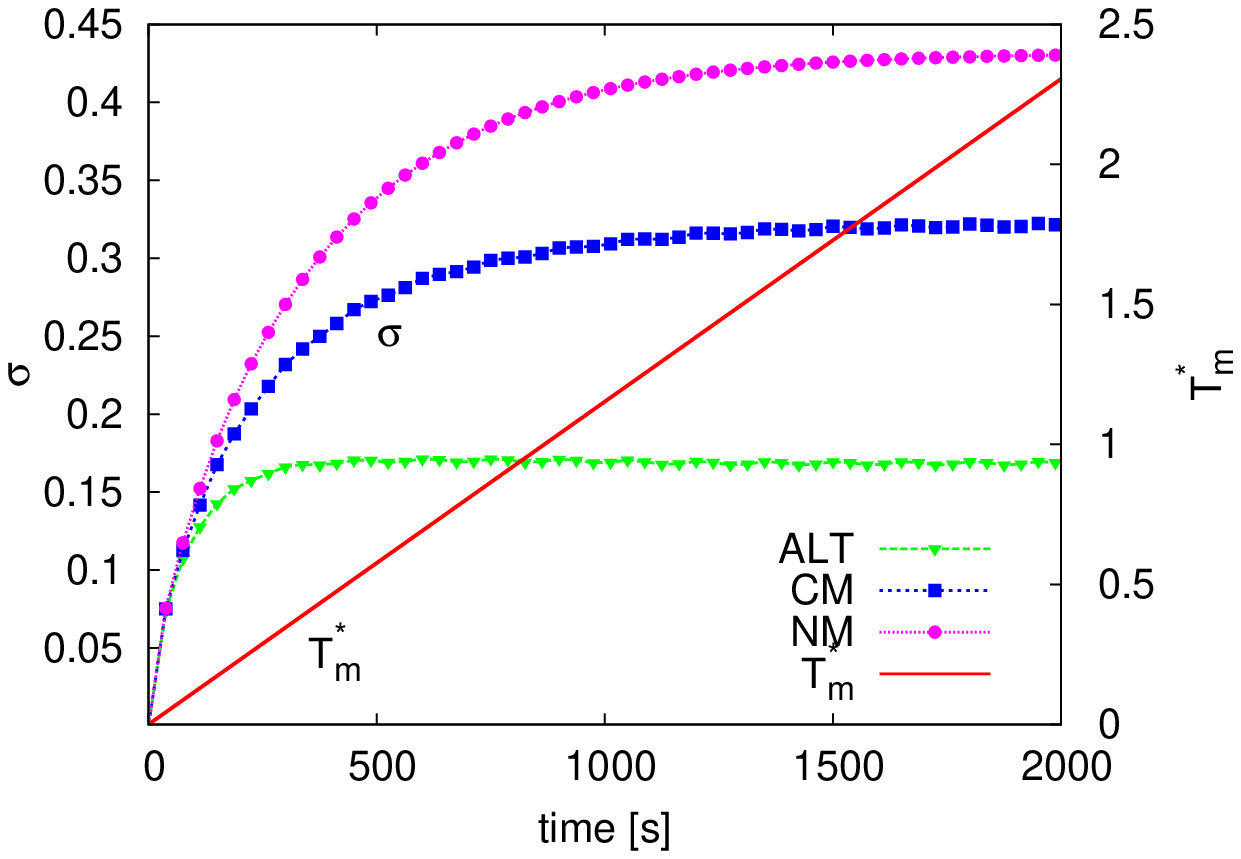}
\includegraphics[width=0.49\textwidth]{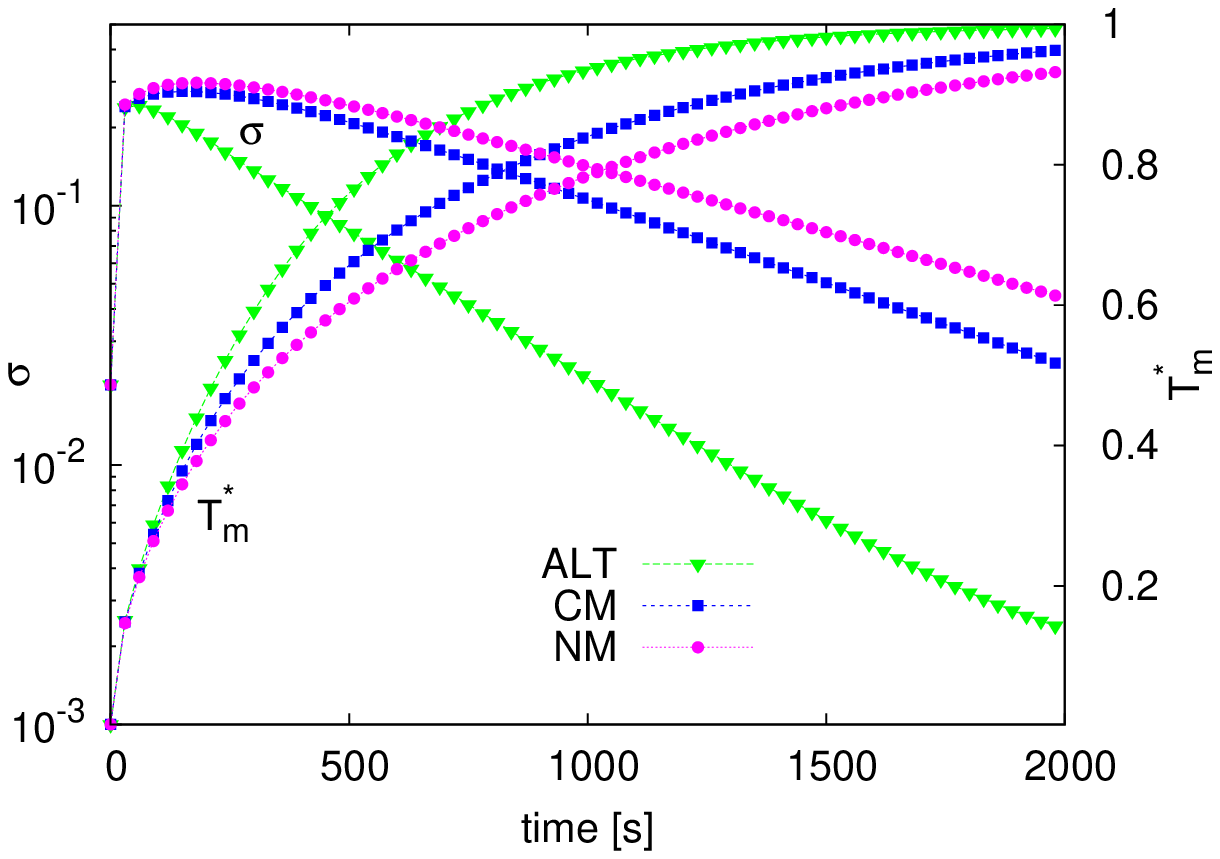}\\
(\textbf{a})\hspace{5cm}(\textbf{b})
\end{center}
 \caption{The temporal evolution of the temperature standard deviation for the non-modulated (NM), continuously modulated (CM) and alternated (ALT) stirring protocols for a modulation period of $30\ s$ and the flow configuration C of Table \ref{tab:stirringconf}. (a) The imposed heat flux. (b) The imposed temperature (note that in this case the $\sigma$ axis is in the logarithmic scale).  }
 \label{fig:sigma_mods}
\end{figure}

The temporal evolutions of the standard deviation $\sigma$ of the temperature scalar and its mean value are shown in Fig. \ref{fig:sigma_mods} for the three stirring protocols, NM, CM and ALT. The results of the imposed heat flux as a boundary condition are compared to those of the imposed temperature for a modulation period of $30\ s$ and for the same flow configuration C of Table~\ref{tab:stirringconf}.

In the first instants ($t \lesssim 100\ s$), the evolutions of $\sigma$ are the same because during this time, the mechanism of heat transfer is only governed by thermal diffusion near the wall boundaries. Afterwards, the evolutions of $\sigma$ differ according to the stirring condition and/or protocol chosen.\\ 
In Fig.~\ref{fig:sigma_mods}(a) (imposed heat flux), each curve of $\sigma$ displays a rapid exponential increase before reaching a plateau value. At the beginning of the mixing process, the temperature fluctuations in the mixer increase by much more rate than the increase in the mean value $T^*_m$, resulting in an increase of $\sigma$, and later, an equilibrium is reached (among stretching, folding and diffusion), indicating that the total variation $T^*-T^*_m$ does not evolve any more over the mixer section (see also Fig.~\ref{fig:temperatures}(a) hereafter). This equilibrium is reached more rapidly for the ALT protocol. This is a major difference when compared to the case of the constant wall temperature boundary condition (Fig.~\ref{fig:sigma_mods}(b)), where we found that $\sigma$ decreases exponentially, indicating a tendency towards complete homogenization. As described earlier, while the evolution of $T^*_m$ in the first case is a straight line with a predefined slope (see section \ref{wallboundary}), the curves obtained in the second case present an asymptotic behavior towards the maximum value 1. The rate of this evolution depends on the efficiency of the mixing protocol.

In Fig.~\ref{fig:sigma_mods}(a) we observe that the lower $\sigma$ plateau value is obtained for the ALT stirring protocol, whereas the highest is obtained for the NM stirring protocol. Also $95$\% of the final value of $\sigma$ is obtained after $250s$ for the ALT stirring protocol against $1500s$ for the NM stirring protocol which represents a speed of homogenization of temperature fluctuations roughly $5$ times faster. An asymptotic thermal regime is then more rapidly achieved for the ALT stirring protocol. This confirms what was observed in Fig.~\ref{fig:snapshot_temp}. For the constant wall temperature boundary condition (Fig.~\ref{fig:sigma_mods}(b)) we have obtained the same result, i.e., the rate of the decrease of $\sigma$ is higher for the ALT stirring. We infer that in order to avoid the persistence of closed streamlines near the walls, it is necessary to alternately move these walls. Thus, the heat transport from the walls can be enhanced. Similarly to the NM stirring protocols, the CM stirring protocols also give rise to closed streamlines in the vicinity of the walls and these closed streamlines prevent the radial transport of the temperature scalar inside the mixer. For the NM and CM cases, close to the rotating boundaries, the radial heat transfer is made only by conduction across the streamlines.

\subsection{Influence of flow configurations and of period length}

\begin{figure}[tbh]
 \begin{center}
 \includegraphics[width=0.7\textwidth]{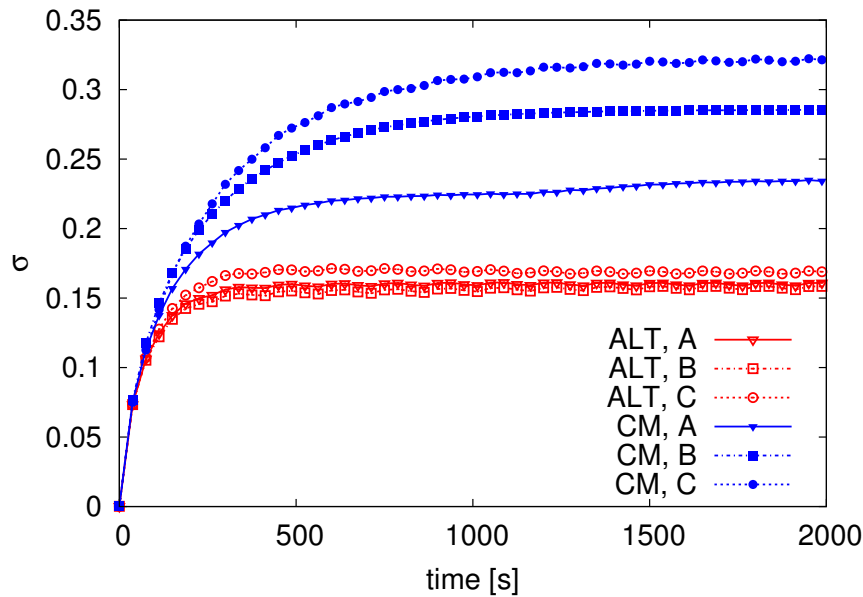}
\end{center}
 \caption{Temporal evolution of temperature standard deviation for the continuously modulated (CM) and alternated (ALT) stirring protocols for a modulation period of $30\ s$ and flow configurations A, B and C of Table \ref{tab:stirringconf}.}
 \label{fig:sigma_stirr}
\end{figure}

Mixing in the two-rod mixer can be conducted in different manners with respect to the direction of rotation of the three walls, as listed in Table \ref{tab:stirringconf}. In Fig.~\ref{fig:sigma_stirr}, the evolutions of $\sigma$ are compared for the cases of the CM and ALT stirring protocols with a modulation period of $30\ s$ and flow configurations A, B and C. Regardless of the flow configuration, the ALT stirring protocol always gives the best result (i.e., lower value of $\sigma$). For this stirring protocol, the values of $\sigma$ are almost insensitive to the choice of the flow configuration. This behavior is also observed for the constant wall boundary condition (see Fig. 13 of reference \citep{elomari2009a}). The CM stirring protocol results, however, depend on the flow configuration, and the best efficiency is obtained with configuration A.

\begin{figure}[tbh]
 \begin{center}
 \includegraphics[width=0.7\textwidth]{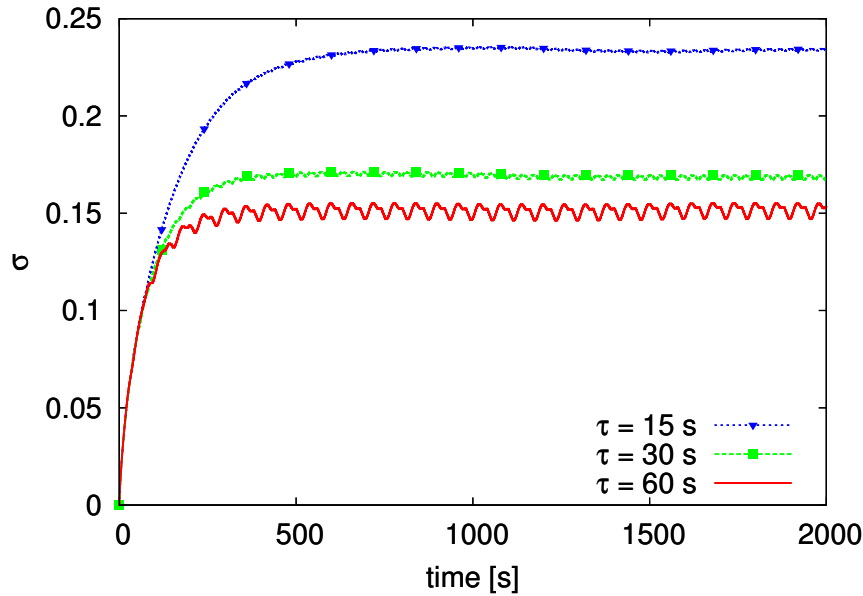} 
\end{center}
 \caption{The temporal evolution of the temperature standard deviation for the ALT stirring protocol and flow configuration C of Table \ref{tab:stirringconf} for three modulation periods: $15\ s$, $30\ s$ and $60\ s$.}
 \label{fig:sigma_periods}
\end{figure}

Another parameter that impacts this mixing process is the length of the modulation period $\tau$ of the angular velocity. Fig. \ref{fig:sigma_periods} shows the influence of the duration of the modulation period on the $\sigma$ value for the ALT stirring protocol and flow configuration C. For this flow configuration, longer duration of the modulation period is related to lower $\sigma$ value as well as better mixing and heat transfer efficiency. Small periodic oscillations are observed on the curves with the same period as the modulation period. Their amplitudes are highest for the longest modulation period (i.e. $60\ s$). An exhaustive study of the impact of the period size was reported in \citep{elomari2010a} (for Dirichlet boundary condition), which corroborates these findings and indicates that there is an optimum of $\tau$ at about $60\ s$ for configuration C and at about $30\ s$ for configuration B.

\subsection{The evolution of extremum quantities}\label{sec:extremums}

\begin{figure}[tbhp]
 \centering
 \begin{center}
  \includegraphics[width=0.49\textwidth]{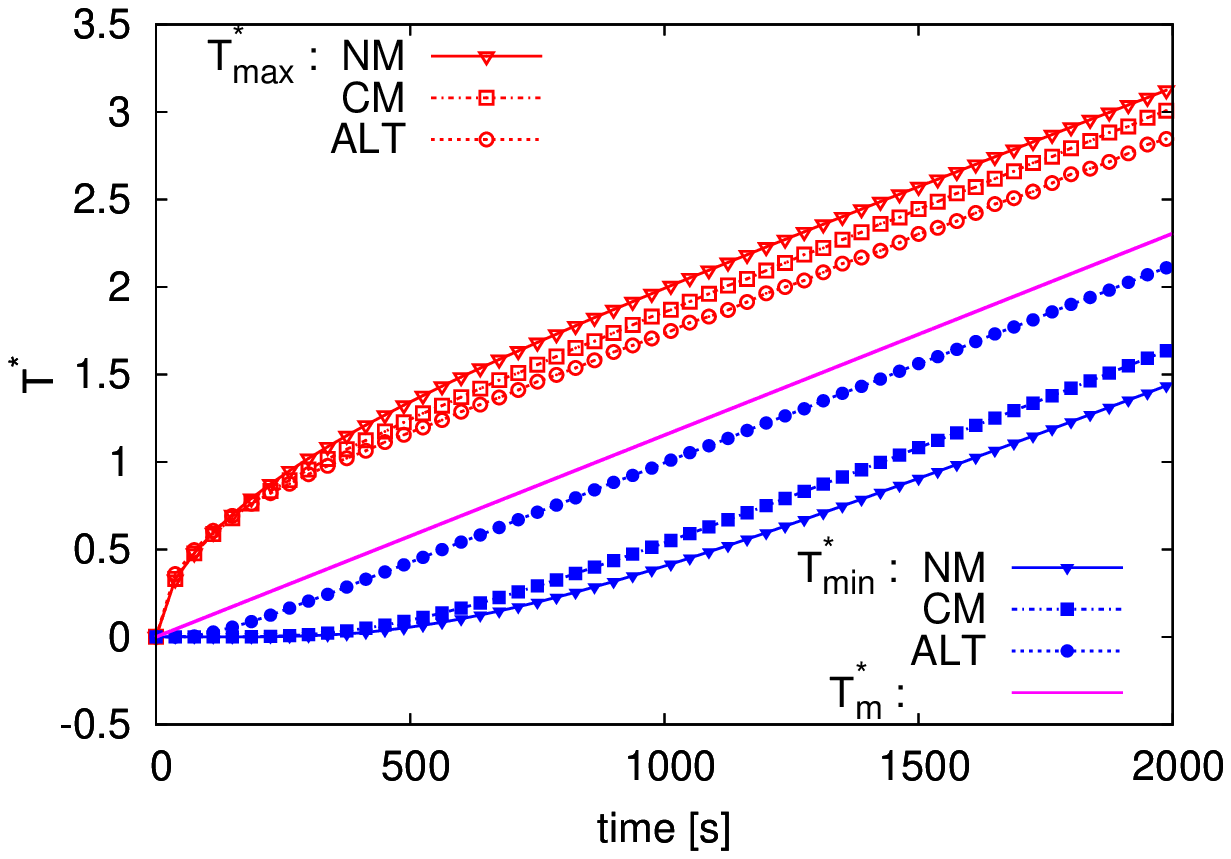}
 \includegraphics[width=0.49\textwidth]{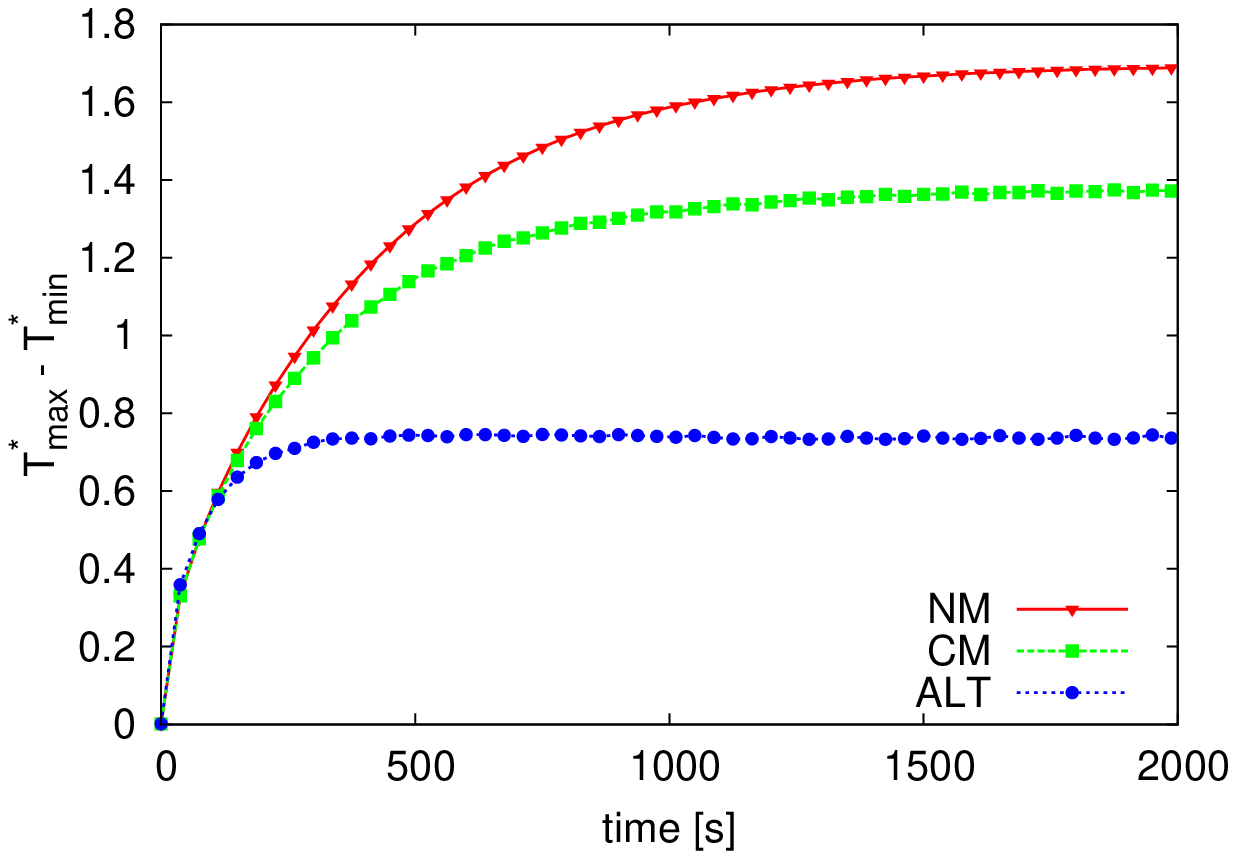}\\
(\textbf{a})\hspace{5cm} (\textbf{b})\\
  \includegraphics[width=0.5\textwidth]{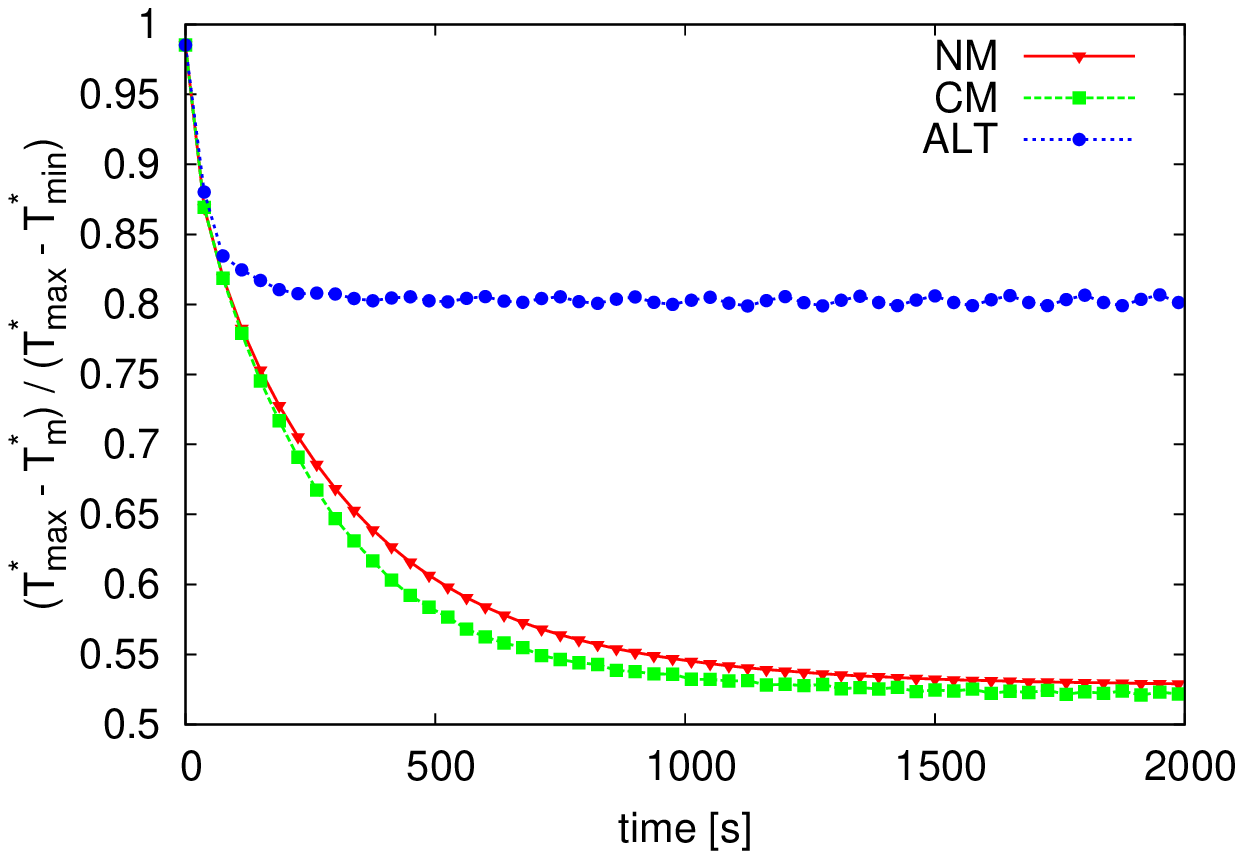}\\(\textbf{c})
\end{center}
 \caption{For the flow configuration C of Table \ref{tab:stirringconf} and the three stirring protocols (NM, CM and ALT): (a) the temporal evolutions of $T^*_{max}$, $T^*_{m}$ and $T^*_{min}$, (b) the temporal evolutions of $T^*_{max} - T^*_{min}$ and (c) the temporal evolutions of $(T^*_{max} - T^*_{m})/(T^*_{max} - T^*_{min})$.}
 \label{fig:temperatures}
\end{figure}

The temporal evolutions of the maximum, mean and minimum temperature scalar values (and also some deduced values) are shown in Fig. \ref{fig:temperatures} for the stirring flow C and the three stirring protocols: NM , CM and ALT.

As mentioned in the section \ref{wallboundary}, we can verify (see Fig. \ref{fig:temperatures}(a)) that for each stirring protocol, the mean fluid temperature evolves linearly with time with the same slope regardless of the flow conditions. Thus, the interesting quantities to follow with time are the maximum and minimum temperature values. We observe that after a transition period, the evolutions of the maximum and minimum temperatures also become linear, and evolve with the same slope as the mean temperature regardless of the stirring protocol. The stirring protocol influences the gaps between the $T^*_{max}$ and $T^*_{min}$ curves and their respective position relative to the $T^*_m$ one.

Hence, it is interesting to examine the temporal evolutions of $T^*_{max} - T^*_{min}$ plotted in Fig.~\ref{fig:temperatures}(b), which shows that they have some shape similarity with those of $\sigma$ (see Fig. \ref{fig:sigma_mods} (a)).
 In accordance with the results given by the temporal evolutions of the standard deviation of the temperature scalar (Fig.~\ref{fig:sigma_mods}(a)), the ALT stirring protocol gives the minimum temperature difference $T^*_{max} - T^*_{min}$ and the NM stirring protocol gives the maximum one. Also, for the ALT stirring protocol $T^*_{min}$ has the closest values to $ T^*_{m}$. We recall that in the case of the fixed temperature boundary condition, the maximum temperature in the mixer is constant and only $T^*_m$ (Fig.~\ref{fig:sigma_mods}(b) ) and the $T^*_{min}$ are increasing.

In the case of heating and mixing with a fixed heat flux, the control of $T^*_{min}$ to prevent the persistence of cold spots is also problematic, and a particular effort has to be made to keep this minimal value as close as possible to the mean value. For this purpose, the last plot of Fig. \ref{fig:temperatures} shows the temporal evolutions of the quantity $\frac {T^*_{max} - T^*_{m}}{T^*_{max} - T^*_{min}}$. With this representation of the results, we can see the relative position of  $ T^*_{m}$ in the interval $[T^*_{max} , T^*_{min}]$, and as before, the alternated stirring protocol always gives the best mixing and heat transfer efficiency. For this case $T^*_{max} - T^*_{m}$ represents $80 \%$ of the total variation $T^*_{max} - T^*_{min}$ ($T^*_{min}$ is close to $T^*_{m}$, see Fig. \ref{fig:temperatures}(a)). On the contrary, for the two other stirring protocols, $T^*_{max} - T^*_{m}$ is around $50 \%$ of the total variation $T^*_{max} - T^*_{min}$.

\subsection{Probability distribution functions of the temperature scalars}

\begin{figure}[tbh]
 \centering
 \begin{center}
  \includegraphics[width=0.7\textwidth]{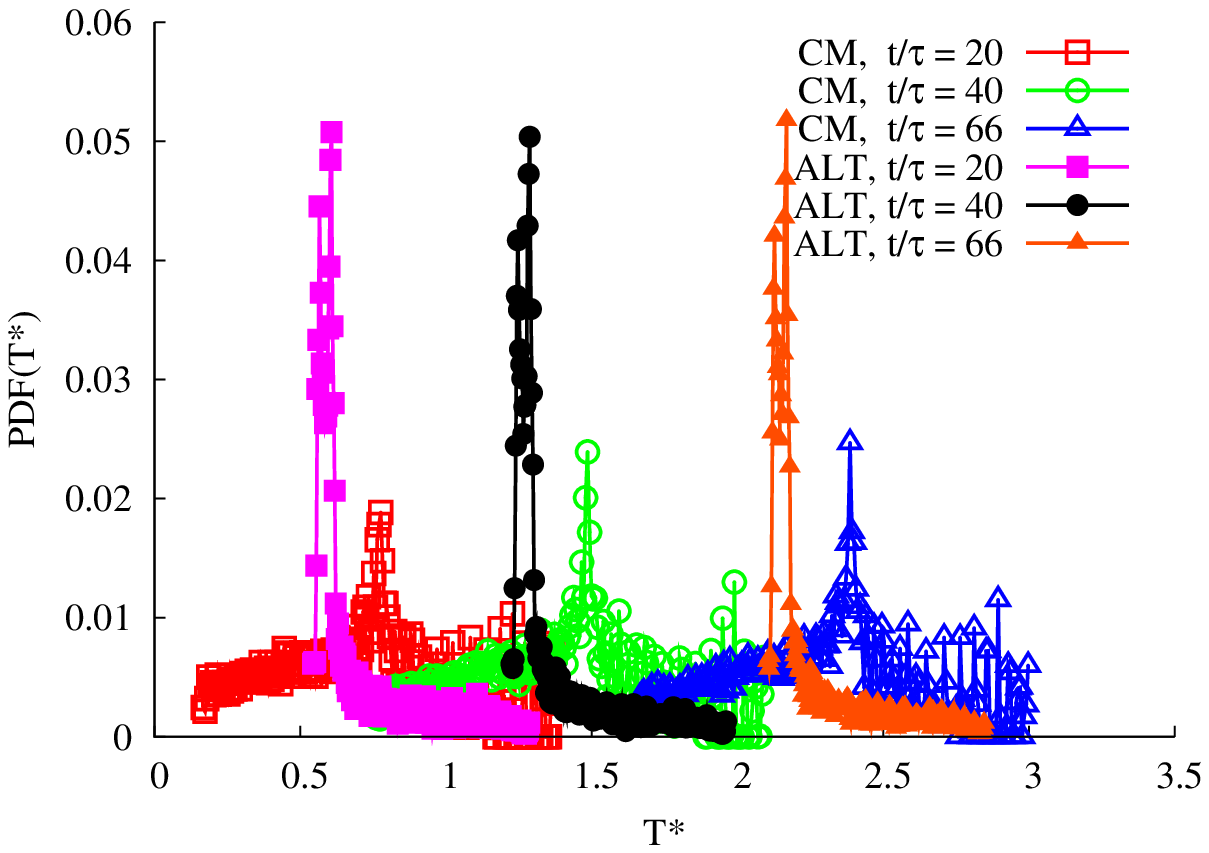}\\(\textbf{a})\\
  \includegraphics[width=0.7\textwidth]{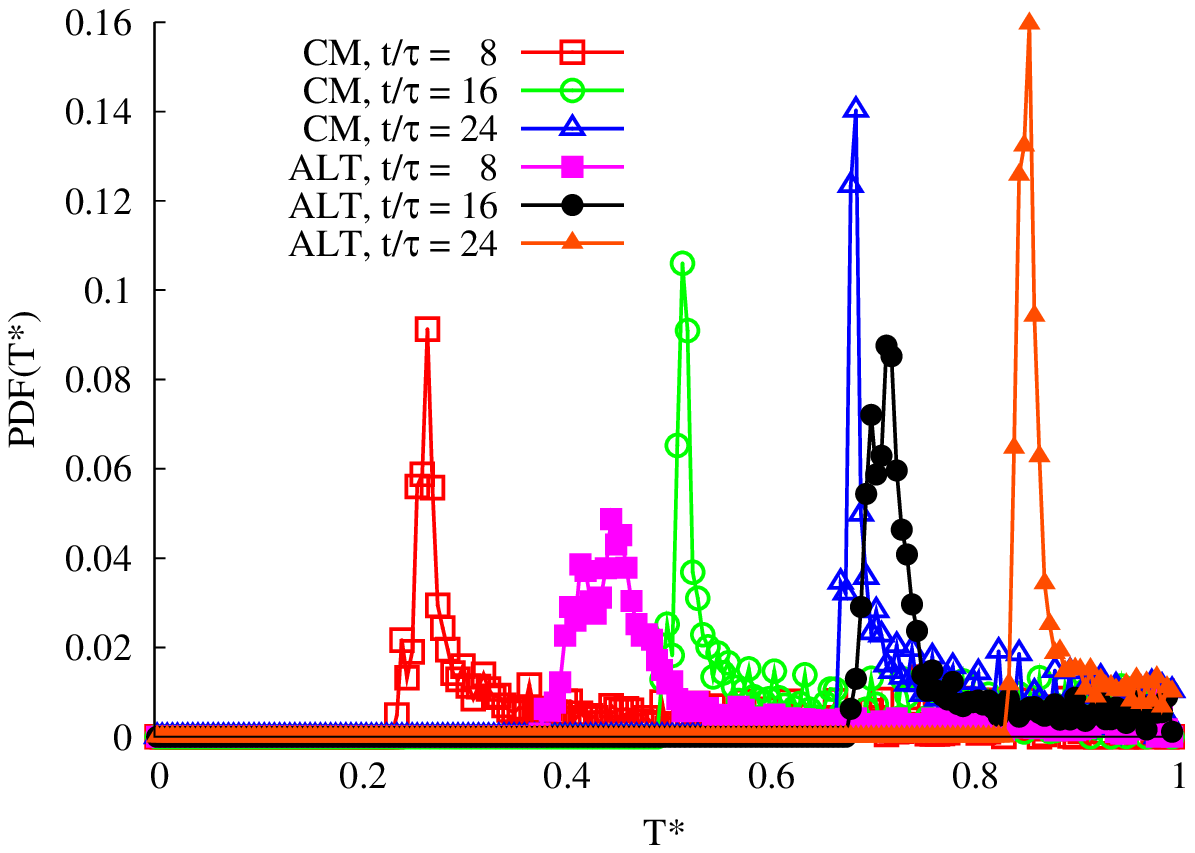}\\(\textbf{b})
\end{center}
 \caption{PDFs of the dimensionless temperature $T^{*}$ at different periodic times for the CM and ALT stirring protocols and flow configuration C ($\tau =30\ s$). (a) The fixed heat flux boundary condition, and (b) the fixed temperature boundary condition.}
 \label{fig:PDF_T}
\end{figure}

In this section, we examine the statistics of the temperature via the analysis of the probability distribution functions (PDFs) of the dimensionless temperature $T^*$ for the whole mixer section filled by the fluid. 

The PDFs of $T^*$ fields are shown in Fig. \ref{fig:PDF_T} for the flow configuration C, at different periodic times (for $\tau =30\ s$) and for the CM and ALT stirring protocols. 
In all of the PDFs, we notice the presence of a significant peak that corresponds to the most probable temperatures in the fluid. In the case of the imposed heat flux (Fig. \ref{fig:PDF_T}(a)), while the PDFs of the CM stirring protocol are almost symmetrical, those of the ALT stirring protocol do not have a left tail. This indicates that the coldest temperatures have disappeared. Furthermore, the ALT protocol shows a higher peak (more exactly a bi-peak and at that particular time of the period) that translates towards the hot temperatures as time evolves. In both cases (CM and ALT), the PDFs recover their forms from period to period. Although the most probable temperature of the CM protocol is higher than that of the ALT protocol, its extended left tail manifests a lack of homogenization.

When a constant temperature is imposed as the boundary condition (Fig. \ref{fig:PDF_T}(b)), both the CM and ALT protocols do not have left tails. The peaks of the ALT protocol are located to the right of the CM peaks, towards the higher temperatures, but with lower amplitudes for the early times. For this type of boundary condition, when the peaks move towards the limit value $T^*=1$ their heights increase to reach the probability value of almost 1. On the contrary, for the imposed heat flux, the PDFs keep the same distribution while translating along the $T^*$ axis, a fact already shown for the evolutions of $T^*_{max} -T^*_{min}$ and $T^*_{max} -T^*_{m}$.

\begin{figure}[tbh]
 \centering
 \begin{center}
\includegraphics[width=0.7\textwidth]{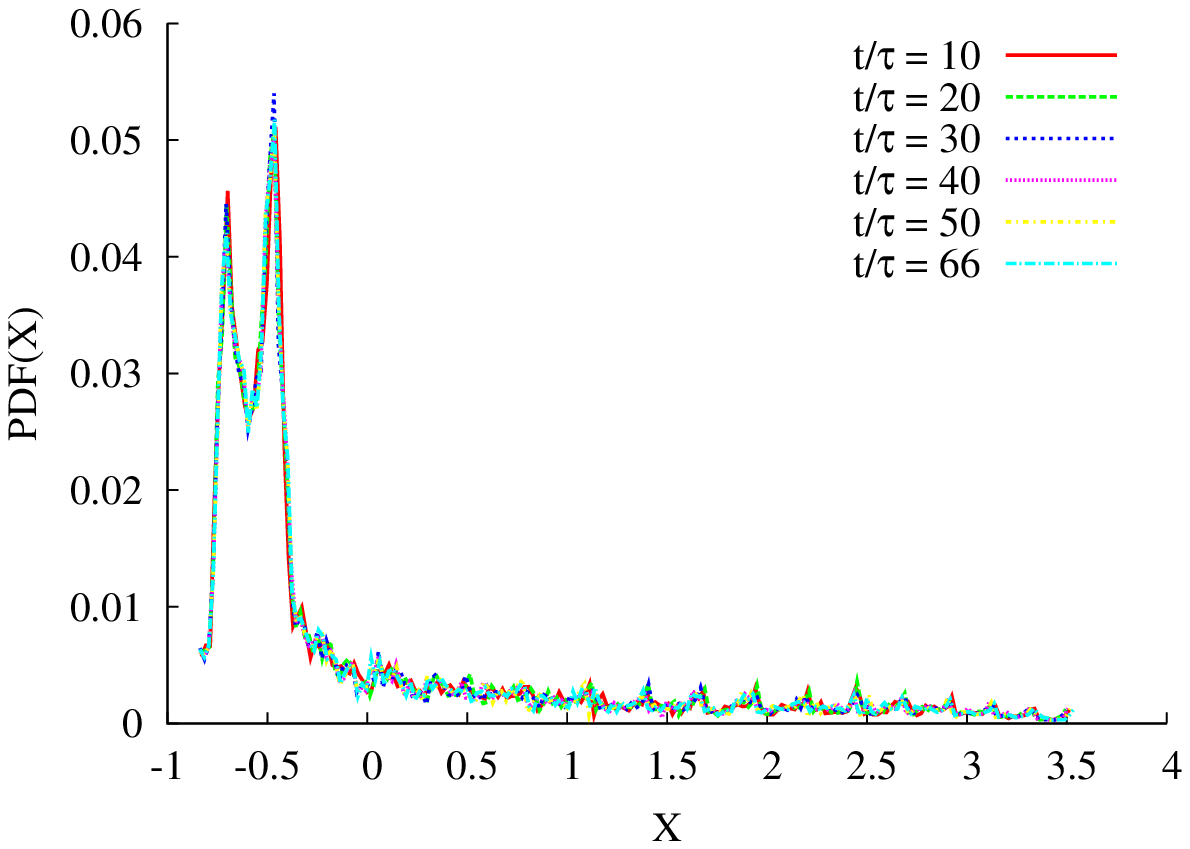}
\end{center}
 \caption{PDFs of the rescaled dimensionless temperature $X$ at different periodic times for the ALT stirring protocols and the flow configuration C ($\tau =30\ s$).}
 \label{fig:PDF_X}
\end{figure}

In Fig. \ref{fig:PDF_X}, are shown for the ALT stirring protocol the PDFs of the rescaled dimensionless temperature:
\begin{equation}
 X = \dfrac{T^*-T^*_m}{\sigma}
 \label{eq:resdimtemp}
\end{equation}  

These PDFs are superimposed when plotted for different times during the mixing process, but at the same phase of the period. This is the signature of a strange eigenmode \citep{elomari2010a,gouillart2008}, which is characterized by the production of persistent (self-similar) patterns in the flow. These patterns arise from a combination of stretching, folding and thermal diffusion. In fact, the temperature distribution evolves within the fluid during a period $\tau$, but it regains its previous form of one period earlier.

%%%%%%%%%%%%%%%%%%%%%%%%%%%%%%%%%%%%%%%%
\subsection{Nusselt number evolutions}

\begin{figure}[tbh]
 \centering
 \begin{center}
  \includegraphics[width=0.7\textwidth]{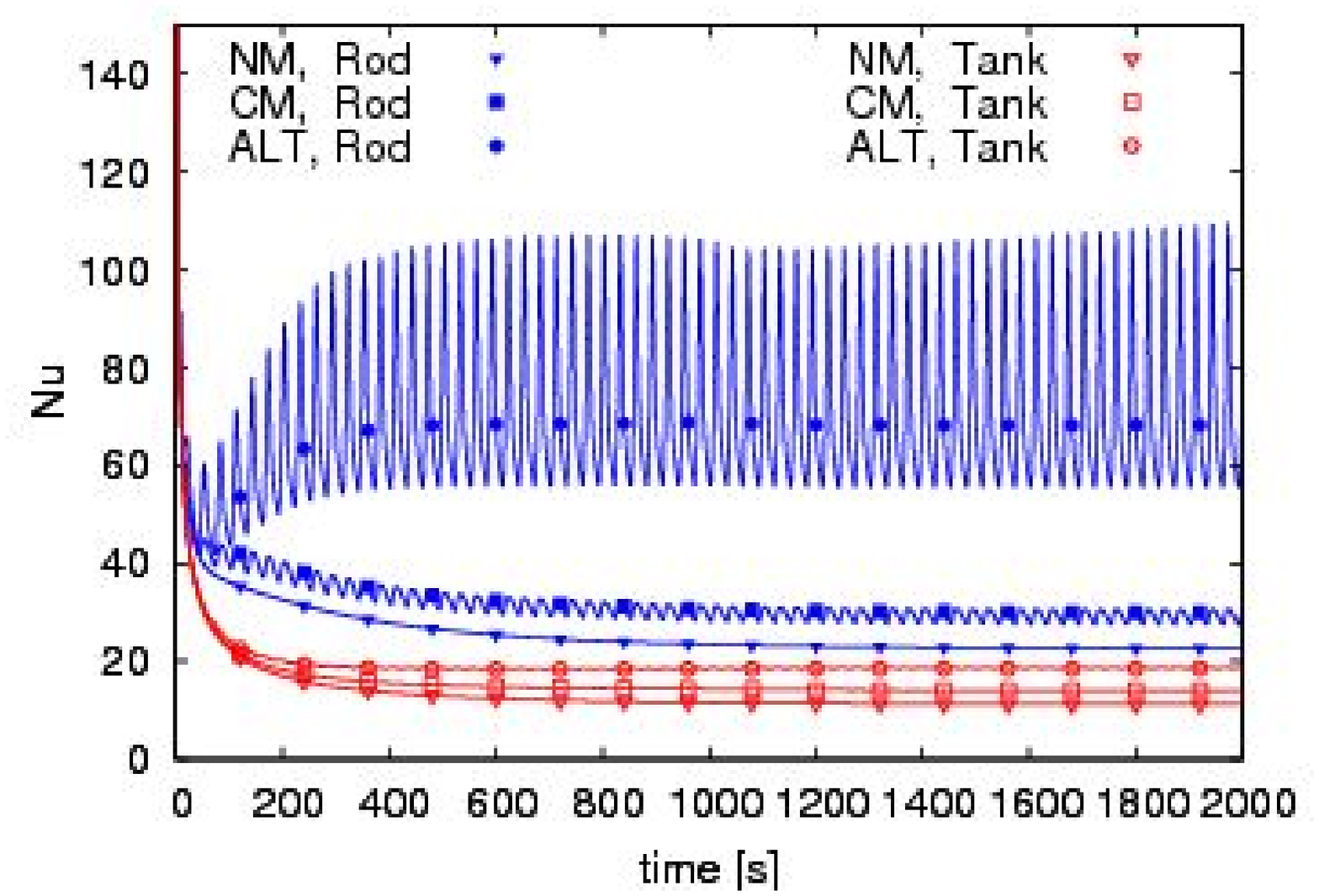}\\(\textbf{a})\\
\includegraphics[width=0.7\textwidth]{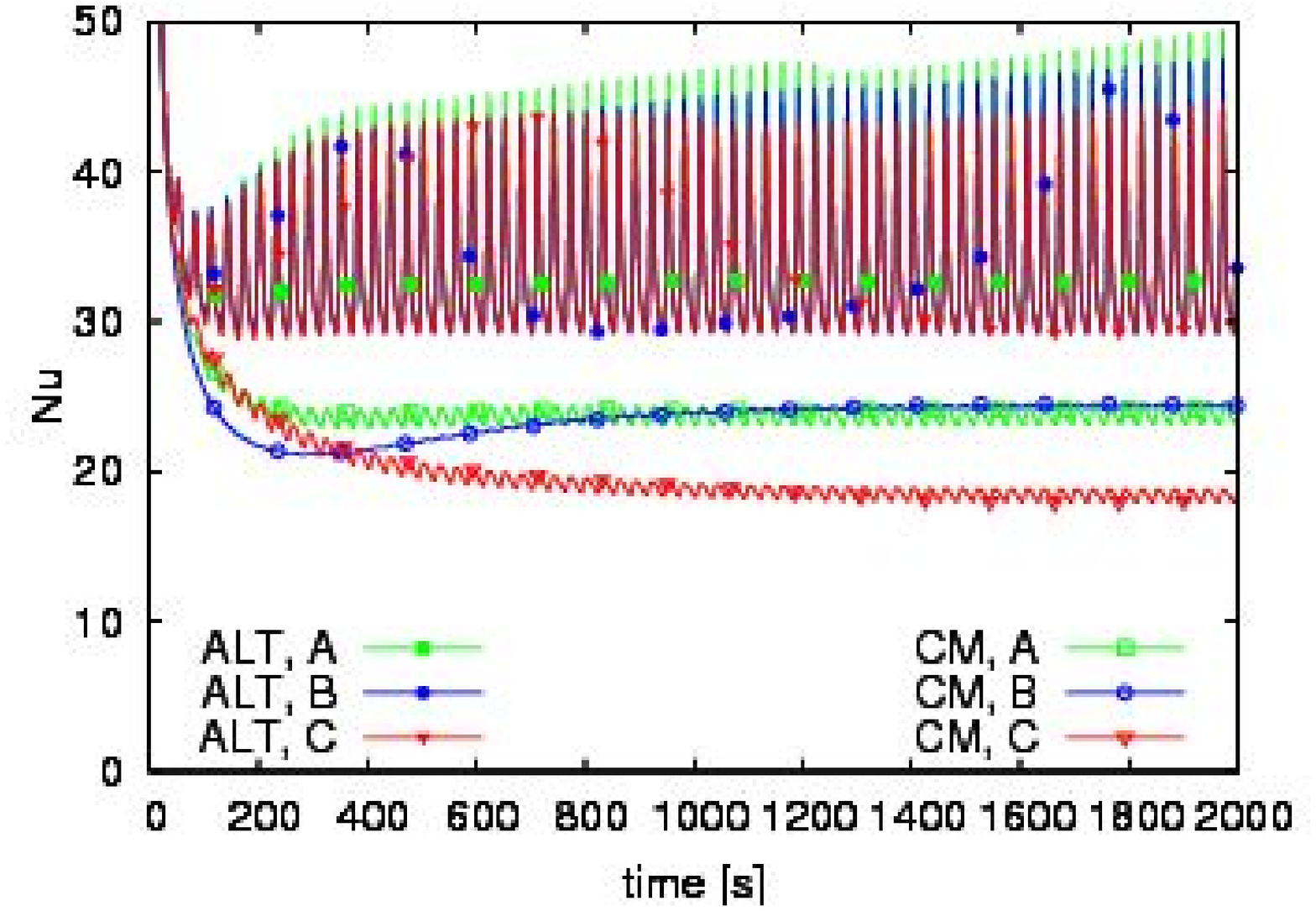}\\(\textbf{b})
\end{center}
 \caption{The Nusselt number evolutions: \textbf{(a)} for the NM, CM and ALT stirring protocols with the rod and tank contributions separated (flow configuration C), and \textbf{(b) }for the flow configurations A, B and C and for the CM and ALT stirring protocols with the rod and tank contributions combined.}
 \label{fig:nusselt}
\end{figure}

The local heat transfer rate is quantified by the calculation of the mean Nusselt number, as indicated in section \ref{mixingindicators}. The temporal evolutions of the Nusselt numbers are given separately for the rods and tank in Fig. \ref{fig:nusselt}(a) and for all of the walls in Fig. \ref{fig:nusselt}(b). At the beginning of the process, the flow is not thermally or hydrodynamically developed, thus the Nusselt number is higher. After a short time, the Nusselt number decreases and reaches an asymptotic state. This asymptotic state corresponds to a constant value for the CM stirring protocol and to a periodic state for the ALT stirring protocol (with a period identical to that of the wall velocity modulation).

Due to this periodic wall displacement, the ALT stirring protocols present effectively large oscillations of the Nusselt number evolutions, and these oscillations are mainly given by the flow behavior around the rods [see Fig. \ref{fig:nusselt}(a)]. When summing the contributions of all the walls [Fig. \ref{fig:nusselt}(b)], the mean value of \textit{Nu} given by the ALT stirring protocol is always greater than that given by the CM protocol (by a ratio of about 1.5), even if the walls have periodic stops in the ALT protocol. It is worth to note that the velocities of all the rotating walls are always greater in the case of CM protocol than of ALT one (Fig. \ref{fig:modulation}), however, the Nusselt number is more important for the latter protocol: the parietal heat transfer is more governed by the patterns and the direction of the flow than by the magnitude of the velocity. 
 Moreover, all of the flow configurations (A, B and C) give comparable results for the ALT protocol, while for the CM protocol, the C configuration is gives a lower \textit{Nu}.

\subsection{Temperature gradient evolutions}

\begin{figure}[tbh]
 \centering
 \begin{center}
 \includegraphics[width=0.7\textwidth]{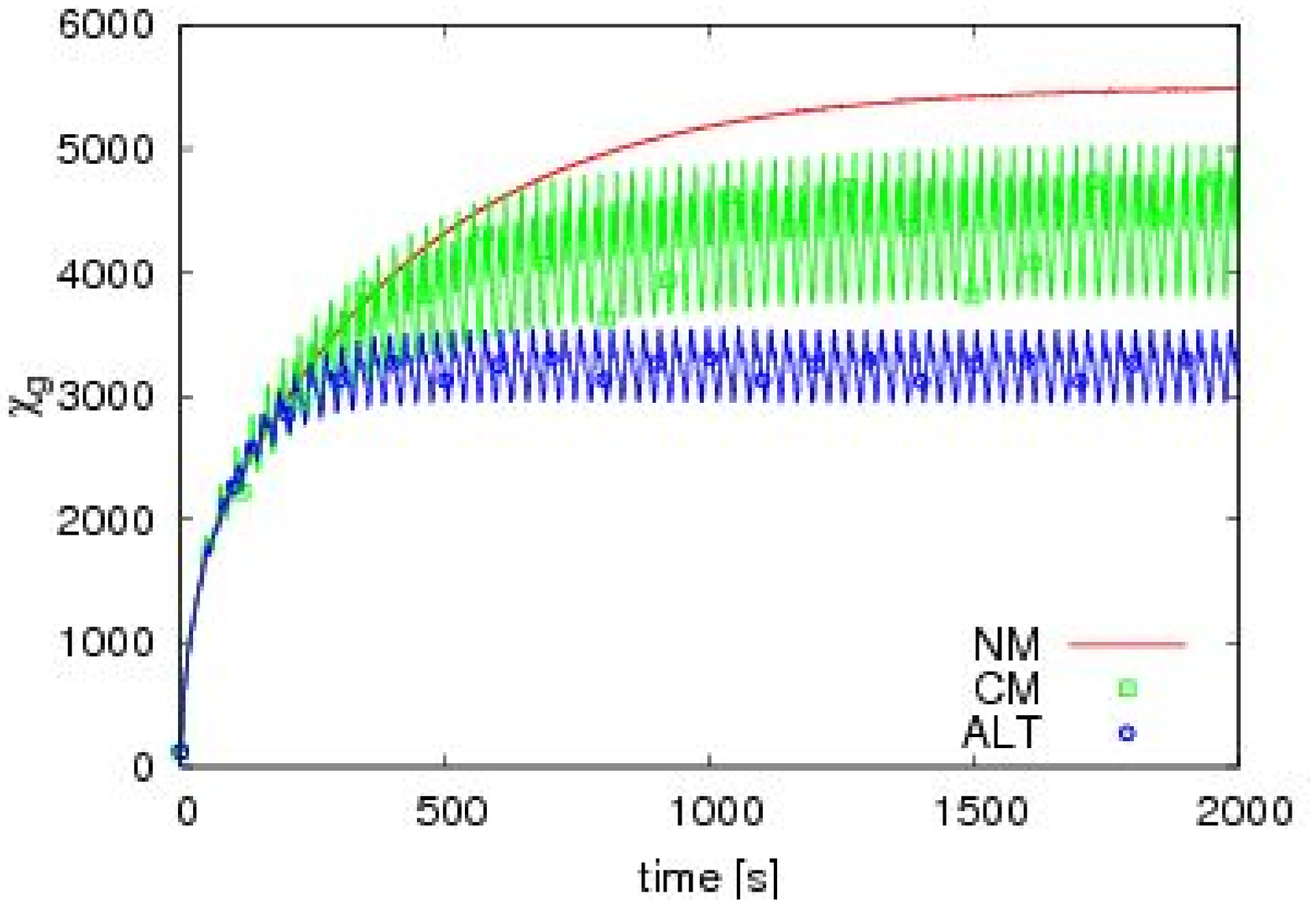}\\\textbf{(a)}\\
 \includegraphics[width=0.7\textwidth]{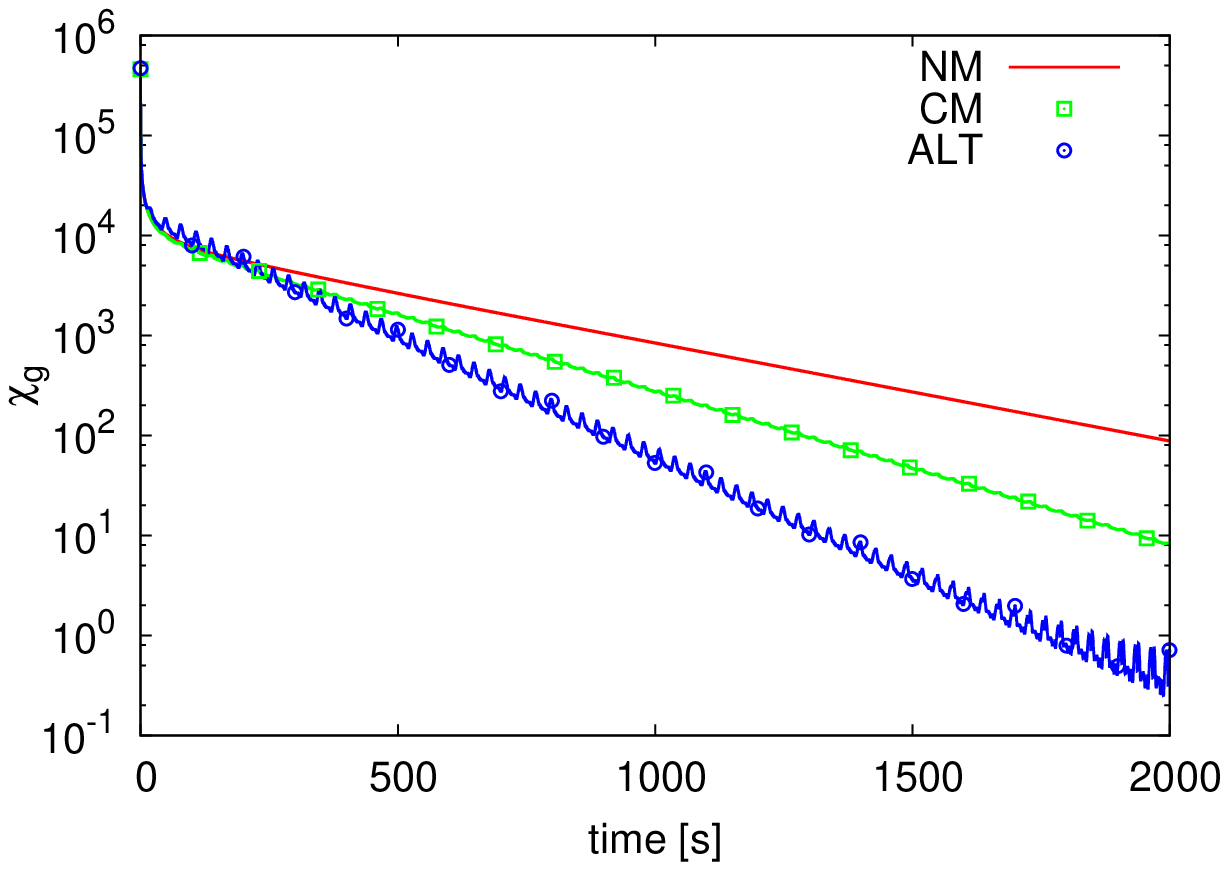}\\\textbf{(b)}
\end{center}
 \caption{The temporal evolutions of the scalar temperature dissipation indicator for the NM, CM and ALT stirring protocols with flow configuration C. \textbf{(a)} The imposed heat flux. \textbf{(b)} The imposed temperature (logarithmic scale).}
 \label{fig:dissipation}
\end{figure}

In Fig.~\ref{fig:dissipation}, we present the evolution of the temperature gradients for the CM and ALT stirring protocols. The temperature scalar dissipation indicator $\chi_g$ defined in section \ref{mixingindicators} is used for this purpose. In the case of the imposed heat flux (Fig.~\ref{fig:dissipation}(a)), the temporal evolutions of $\chi_g$ are similar to those of the temperature standard deviation for the same stirring protocols (see Fig.~\ref{fig:sigma_mods}(a)): after a first phase of gradient creation, its value stabilizes. However, the CM and ALT stirring protocols present large oscillations around their asymptotic value. The gradients of the temperature scalar do not vanish in time as is the case for mixing with a constant wall temperature boundary condition (Fig.~\ref{fig:dissipation}(b)). In the latter case, the decrease of the temperature gradients is exponential, and again its rate is steeper for the ALT protocol.

\section{Conclusion}
\label{conclusion}
We have investigated, by numerical simulations, the coupled mixing and heating performances induced by chaotic advection in a 2D two-rod mixer for a constant wall heat flux boundary condition. Three different stirring protocols were compared: a non-modulated, a continuously modulated and an alternated (non-continuous) one for different flow configurations (depending on the respective direction of the rotation of the rod and tank walls). 
In order to study the thermal mixing enhancement mechanism within the fluids, different mixing indicators and statistical tools were used. 
According to the wall boundary condition considered (constant wall heat flux), the following main conclusions can be made based on the results obtained:
\begin{itemize}
\item[$\bullet$] The most effective stirring protocol for thermal mixing is the alternated one (a conclusion also obtained for the constant wall temperature boundary condition \citep{elomari2010a}).
\item[$\bullet$] For this alternated stirring protocol, the difference in thermal mixing induced by the choice of the flow configuration (A, B or C) is not very sensitive, contrary to the case of the continuously modulated stirring protocol.
\item[$\bullet$] The alternated stirring protocol seems to quickly suppress the coldest temperature during the mixing process, which is again not the case for the continuously modulated stirring protocol.
\item[$\bullet$] A great enhancement of heat transfer is thus obtained for the alternated stirring protocol, confirmed by the Nusselt number evolutions. Despite the oscillations observed for the Nusselt number and temperature gradients evolutions, the homogeneity in the temperature during the thermal chaotic mixing is conserved (i.e. constant temperature standard deviation). 

\end{itemize}

We also showed the major differences between the two modes of parietal heating (fixed temperature or fixed heat flux) in the context of chaotic mixing.  These differences can be summerized in Tab. \ref{tab:heat_modes}.

\begin{table}[htbp]
 \begin{center}
\begin{tabular}[b]{p{0.1\textwidth} p{0.4\textwidth} p{0.4\textwidth} }
%\cline{2-3}
& \textbf{Fixed temperature} & \textbf{Fixed heat flux}\\ \hline
\centering Heat absorbed & The amount of absorbed heat by the fluid (and $T_m$) depends on the flow. &  The amount of absorbed heat if fixed.\\ \\ %\hline
\centering $T_m$ & The evolution of $T_m$ is asymptotic. &  The evolution of $T_m$ is linear and continuously evolving.\\ \\ %\hline
\centering $\sigma$ & A high degree of homogenization (very low $\sigma$) can be achieved as the mixing process is continued. $\sigma$ decreases exponentially with time. & The degree of homogenization settles at a fixed value.\\ \\%\hline
\centering Asymtotic thermal regime & For $T_m$ or $T^*_{m}$ &   For $\sigma$,    $T^*_{max}-T^*_{min}$ or $T_{max}^*-T^*_{m}$, $\chi_g$.\\ \\%\hline
\centering Mixing efficiency & The efficiency of the mixing protocol impacts the rate of homogenization and the decay rate of the temperature gradient $\chi_g$. & The efficiency of the mixing protocol impacts the difference between the extrema of the temperatures, as well as $\sigma$ and $\chi_g$. All these values remain constant as the mixing proceeds.\\\hline
\end{tabular}
\caption{Major differences between the fixed temperature and the fixed heat flux modes of parietal heating.}
\label{tab:heat_modes}
\end{center}
\end{table}

% \begin{center}
% \hspace{0.05\textwidth}\parbox{0.4\textwidth}{\textbf{Fixed temperature} }\hspace{0.05\textwidth}\parbox{0.4\textwidth}{\textbf{Fixed heat flux}}\\
%  \rule{0.5\textwidth}{1pt}\\
% \parbox{0.45\textwidth}{The amount of absorbed heat by the fluid (and $T_m$) depends on the flow.} \hspace{0.05\textwidth}
% \parbox{0.45\textwidth}{The amount of absorbed heat if fixed.}\\
%  \rule{0.5\textwidth}{1pt}\\
% \parbox{0.45\textwidth}{The evolution of $T_m$ is asymptotic.} \hspace{0.05\textwidth} \parbox{0.45\textwidth}{The evolution of $T_m$ is linear and continuously evolving.}
% \end{center}

The mixing and heating using a prescribed heat flux always gives a non-homogenized fluid temperature. On the other hand, heating with a fixed temperature suffers from the asymptotic evolution of the quantity of heat transmitted to the fluid. In an industrial process, the heating of a highly viscous fluid can be first started with a fixed flux to benefit from a high rate of energy supply. An alternated protocol will ensure a low gradient in the fluid and will help to maintain the maximal temperature below the fluid damage temperature. When the target temperature is approached, it can be fixed at the mixer walls to finish the mixing with the suitable degree of homogenization.

Further studies will explore the effect of the temperature-dependence of the physical properties on the efficiency of the thermal chaotic mixing.

\bibliographystyle{elsarticle-num-names}
%\bibliography{../KEO_global_biblio}

\begin{thebibliography}{34}
\providecommand{\natexlab}[1]{#1}
\providecommand{\url}[1]{\texttt{#1}}
\providecommand{\urlprefix}{URL }
\expandafter\ifx\csname urlstyle\endcsname\relax
  \providecommand{\doi}[1]{doi:\discretionary{}{}{}#1}\else
  \providecommand{\doi}[1]{doi:\discretionary{}{}{}\begingroup
  \urlstyle{rm}\url{#1}\endgroup}\fi
\providecommand{\bibinfo}[2]{#2}

\bibitem[{Ottino(1989)}]{ottino1989}
\bibinfo{author}{J.~M. Ottino}, \bibinfo{title}{{The kinematics of mixing:
  stretching, chaos, and transport}}, \bibinfo{publisher}{Cambridge University
  Press}, \bibinfo{year}{1989}.

\bibitem[{Aref(2002)}]{aref2002}
\bibinfo{author}{H.~Aref}, \bibinfo{title}{{The development of chaotic
  advection}}, \bibinfo{journal}{Physics of Fluids} \bibinfo{volume}{14}
  (\bibinfo{year}{2002}) \bibinfo{pages}{1315--1325}.

\bibitem[{Jana and Sau(2004)}]{jana2004}
\bibinfo{author}{S.~C. Jana}, \bibinfo{author}{M.~Sau},
  \bibinfo{title}{{Effects of viscosity ratio and composition on development of
  morphology in chaotic mixing of polymers}}, \bibinfo{journal}{Polymer}
  \bibinfo{volume}{45}~(\bibinfo{number}{5}) (\bibinfo{year}{2004})
  \bibinfo{pages}{1665--1678}.

\bibitem[{Metcalfe and Lester(2009)}]{metcalfe2009}
\bibinfo{author}{G.~Metcalfe}, \bibinfo{author}{D.~Lester},
  \bibinfo{title}{Mixing and heat transfer of highly viscous food products with
  a continuous chaotic duct flow}, \bibinfo{journal}{Journal of Food
  Engineering} \bibinfo{volume}{95}~(\bibinfo{number}{1})
  (\bibinfo{year}{2009}) \bibinfo{pages}{21--29}.

\bibitem[{Caubet et~al.(2010)Caubet, Le~Guer, Grassl, El~Omari, and
  Normandin}]{caubet2010}
\bibinfo{author}{S.~Caubet}, \bibinfo{author}{Y.~Le~Guer},
  \bibinfo{author}{B.~Grassl}, \bibinfo{author}{K.~El~Omari},
  \bibinfo{author}{E.~Normandin}, \bibinfo{title}{A low-energy emulsification
  batch mixer for concentrated oil-in-water emulsions}, \bibinfo{journal}{AIChE
  Journal, In press.} \doi{\bibinfo{doi}{10.1002/aic.12253}}, \bibinfo{note}{{
  DOI: 10.1002/aic.12253}}.

\bibitem[{Chang and Sen(1994)}]{chang1994}
\bibinfo{author}{H.~C. Chang}, \bibinfo{author}{M.~Sen},
  \bibinfo{title}{{Application of chaotic advection to heat transfer}},
  \bibinfo{journal}{Chaos, Solitons and Fractals}
  \bibinfo{volume}{4}~(\bibinfo{number}{6}) (\bibinfo{year}{1994})
  \bibinfo{pages}{955--976}.

\bibitem[{Acharya et~al.(2001)Acharya, Sen, and Chang}]{acharya2001}
\bibinfo{author}{N.~Acharya}, \bibinfo{author}{M.~Sen}, \bibinfo{author}{H.~C.
  Chang}, \bibinfo{title}{{Analysis of heat transfer enhancement in coiled-tube
  heat exchangers}}, \bibinfo{journal}{International Journal of Heat and Mass
  Transfer} \bibinfo{volume}{44}~(\bibinfo{number}{17}) (\bibinfo{year}{2001})
  \bibinfo{pages}{3189--3199}.

\bibitem[{Peerhossaini et~al.(1993)Peerhossaini, Castelain, and
  Le~Guer}]{peerhossaini1993}
\bibinfo{author}{H.~Peerhossaini}, \bibinfo{author}{C.~Castelain},
  \bibinfo{author}{Y.~Le~Guer}, \bibinfo{title}{{Heat exchanger design based on
  chaotic advection}}, \bibinfo{journal}{Experimental thermal and fluid
  science} \bibinfo{volume}{7}~(\bibinfo{number}{4}) (\bibinfo{year}{1993})
  \bibinfo{pages}{333--344}.

\bibitem[{Mokrani et~al.(1997)Mokrani, Castelain, and
  Peerhossaini}]{mokrani1997}
\bibinfo{author}{A.~Mokrani}, \bibinfo{author}{C.~Castelain},
  \bibinfo{author}{H.~Peerhossaini}, \bibinfo{title}{{The effects of chaotic
  advection on heat transfer}}, \bibinfo{journal}{International Journal of Heat
  and Mass Transfer} \bibinfo{volume}{40}~(\bibinfo{number}{13})
  (\bibinfo{year}{1997}) \bibinfo{pages}{3089--3104}.

\bibitem[{Gouillart et~al.(2008)Gouillart, Dauchot, Dubrulle, Roux, and
  Thiffeault}]{gouillart2008}
\bibinfo{author}{E.~Gouillart}, \bibinfo{author}{O.~Dauchot},
  \bibinfo{author}{B.~Dubrulle}, \bibinfo{author}{S.~Roux},
  \bibinfo{author}{J.~L. Thiffeault}, \bibinfo{title}{{Slow decay of
  concentration variance due to no-slip walls in chaotic mixing}},
  \bibinfo{journal}{Physical Review E}
  \bibinfo{volume}{78}~(\bibinfo{number}{2}) (\bibinfo{year}{2008})
  \bibinfo{pages}{026211}.

\bibitem[{Gouillart et~al.(2010)Gouillart, Thiffeault, and
  Dauchot}]{gouillart2010}
\bibinfo{author}{E.~Gouillart}, \bibinfo{author}{J.~L. Thiffeault},
  \bibinfo{author}{O.~Dauchot}, \bibinfo{title}{{Rotation shields chaotic
  mixing regions from no-slip walls}}, \bibinfo{journal}{Physical Review
  Letters} \bibinfo{volume}{104} (\bibinfo{year}{2010})
  \bibinfo{pages}{204502}.

\bibitem[{El~Omari and Le~Guer(2010{\natexlab{a}})}]{elomari2010a}
\bibinfo{author}{K.~El~Omari}, \bibinfo{author}{Y.~Le~Guer},
  \bibinfo{title}{Alternate rotating walls for thermal chaotic mixing},
  \bibinfo{journal}{International Journal of Heat and Mass Transfer}
  \bibinfo{volume}{53} (\bibinfo{year}{2010}{\natexlab{a}})
  \bibinfo{pages}{123--134}.

\bibitem[{Ghosh et~al.(1992)Ghosh, Chang, and Sen}]{ghosh1992}
\bibinfo{author}{S.~Ghosh}, \bibinfo{author}{H.~C. Chang},
  \bibinfo{author}{M.~Sen}, \bibinfo{title}{{Heat-transfer enhancement due to
  slender recirculation and chaotic transport between counter-rotating
  eccentric cylinders}}, \bibinfo{journal}{Journal of Fluid Mechanics}
  \bibinfo{volume}{238} (\bibinfo{year}{1992}) \bibinfo{pages}{119--154}.

\bibitem[{Ganesan et~al.(1997)Ganesan, Bryden, and Brenner}]{ganesan1997}
\bibinfo{author}{V.~Ganesan}, \bibinfo{author}{M.~D. Bryden},
  \bibinfo{author}{H.~Brenner}, \bibinfo{title}{{Chaotic heat transfer
  enhancement in rotating eccentric annular-flow systems}},
  \bibinfo{journal}{Physics of Fluids} \bibinfo{volume}{9}
  (\bibinfo{year}{1997}) \bibinfo{pages}{1296--1306}.

\bibitem[{Lefevre et~al.(2003)Lefevre, Mota, Rodrigo, and
  Saatdjian}]{lefevre2003}
\bibinfo{author}{A.~Lefevre}, \bibinfo{author}{J.~P.~B. Mota},
  \bibinfo{author}{A.~J.~S. Rodrigo}, \bibinfo{author}{E.~Saatdjian},
  \bibinfo{title}{{Chaotic advection and heat transfer enhancement in Stokes
  flows}}, \bibinfo{journal}{International Journal of Heat and Fluid Flow}
  \bibinfo{volume}{24}~(\bibinfo{number}{3}) (\bibinfo{year}{2003})
  \bibinfo{pages}{310--321}.

\bibitem[{Mota et~al.(2007)Mota, Rodrigo, and Saatdjian}]{mota2007}
\bibinfo{author}{J.~P.~B. Mota}, \bibinfo{author}{A.~J.~S. Rodrigo},
  \bibinfo{author}{E.~Saatdjian}, \bibinfo{title}{{Optimization of
  heat-transfer rate into time-periodic two-dimensional Stokes flows}},
  \bibinfo{journal}{International journal for numerical methods in fluids}
  \bibinfo{volume}{53}~(\bibinfo{number}{6}) (\bibinfo{year}{2007})
  \bibinfo{pages}{915--931}.

\bibitem[{Price et~al.(2003)Price, Mullin, and Kobine}]{price2003}
\bibinfo{author}{T.~J. Price}, \bibinfo{author}{T.~Mullin},
  \bibinfo{author}{J.~J. Kobine}, \bibinfo{title}{{Numerical and experimental
  characterization of a family of two-roll-mill flows}},
  \bibinfo{journal}{Royal Society of London, Proceedings Series A}
  \bibinfo{volume}{459} (\bibinfo{year}{2003}) \bibinfo{pages}{117--135}.

\bibitem[{Young et~al.(2007)Young, Chiu, and Fan}]{young2007}
\bibinfo{author}{D.~L. Young}, \bibinfo{author}{C.~L. Chiu},
  \bibinfo{author}{C.~M. Fan}, \bibinfo{title}{{A hybrid
  Cartesian/immersed-boundary finite-element method for simulating heat and
  flow patterns in a two-roll mill}}, \bibinfo{journal}{Numerical Heat Transfer
  Part B: Fundamentals} \bibinfo{volume}{51}~(\bibinfo{number}{3-4})
  (\bibinfo{year}{2007}) \bibinfo{pages}{251--274}.

\bibitem[{Chiu et~al.(2009)Chiu, Fan, and Young}]{chiu2009}
\bibinfo{author}{C.~Chiu}, \bibinfo{author}{C.~Fan},
  \bibinfo{author}{D.~Young}, \bibinfo{title}{3D hybrid
  Cartesian/immersed-boundary finite-element analysis of heat and flow patterns
  in a two-roll mill}, \bibinfo{journal}{International Journal of Heat and Mass
  Transfer} \bibinfo{volume}{52}~(\bibinfo{number}{7-8}) (\bibinfo{year}{2009})
  \bibinfo{pages}{1677--1689}.

\bibitem[{Jana et~al.(1994)Jana, Metcalfe, and Ottino}]{jana94}
\bibinfo{author}{S.~C. Jana}, \bibinfo{author}{G.~Metcalfe},
  \bibinfo{author}{J.~M. Ottino}, \bibinfo{title}{Experimental and
  computational studies of mixing in complex Stokes flows: the vortex mixing
  flow and multicellular cavity flows}, \bibinfo{journal}{Journal of Fluid
  Mechanics} \bibinfo{volume}{269} (\bibinfo{year}{1994})
  \bibinfo{pages}{199--246}.

\bibitem[{El~Omari and Le~Guer(2009)}]{elomari2009a}
\bibinfo{author}{K.~El~Omari}, \bibinfo{author}{Y.~Le~Guer}, \bibinfo{title}{{A
  numerical study of thermal chaotic mixing in a two rod rotating mixer}},
  \bibinfo{journal}{Computational Thermal Science} \bibinfo{volume}{1}
  (\bibinfo{year}{2009}) \bibinfo{pages}{55--73}.

\bibitem[{El~Omari and Le~Guer(2010{\natexlab{b}})}]{elomari2010b}
\bibinfo{author}{K.~El~Omari}, \bibinfo{author}{Y.~Le~Guer},
  \bibinfo{title}{Thermal chaotic mixing of power law fluids in a mixer with
  alternately-rotating walls.}, \bibinfo{journal}{J. of Non-Newtonian Fluid
  Mechanics} \bibinfo{volume}{165} (\bibinfo{year}{2010}{\natexlab{b}})
  \bibinfo{pages}{641--651}.

\bibitem[{Geuzaine and Remacle(2009)}]{geuzaine2009}
\bibinfo{author}{C.~Geuzaine}, \bibinfo{author}{J.~F. Remacle},
  \bibinfo{title}{{Gmsh: a three-dimensional finite element mesh generator with
  built-in pre-and post-processing facilities}},
  \bibinfo{journal}{International Journal for Numerical Methods in Engineering}
  \bibinfo{volume}{79}~(\bibinfo{number}{11}) (\bibinfo{year}{2009})
  \bibinfo{pages}{1309--1331}.

\bibitem[{Jasak(1996)}]{jasak1996}
\bibinfo{author}{H.~Jasak}, \bibinfo{title}{{Error analysis and estimation for
  the finite volume method with applications to fluid flows}}, Ph.D. thesis,
  \bibinfo{school}{University of London}, \bibinfo{year}{1996}.

\bibitem[{Ferziger and Peri{\'c}(2002)}]{ferziger2002}
\bibinfo{author}{J.~H. Ferziger}, \bibinfo{author}{M.~Peri{\'c}},
  \bibinfo{title}{{Computational methods for fluid dynamics}},
  \bibinfo{publisher}{Springer New York}, \bibinfo{year}{2002}.

\bibitem[{Alves et~al.(2003)Alves, Oliveira, and Pinho}]{alves2003}
\bibinfo{author}{M.~A. Alves}, \bibinfo{author}{P.~J. Oliveira},
  \bibinfo{author}{F.~T. Pinho}, \bibinfo{title}{{A convergent and universally
  bounded interpolation scheme for the treatment of advection}},
  \bibinfo{journal}{International Journal for Numerical Methods in Fluids}
  \bibinfo{volume}{41} (\bibinfo{year}{2003}) \bibinfo{pages}{47--75}.

\bibitem[{Ng et~al.(2007)Ng, Yusoff, and Ng}]{ng2007}
\bibinfo{author}{K.~C. Ng}, \bibinfo{author}{M.~Z. Yusoff},
  \bibinfo{author}{E.~Y.~K. Ng}, \bibinfo{title}{{Higher-order bounded
  differencing schemes for compressible and incompressible flows}},
  \bibinfo{journal}{International Journal for Numerical Methods in Fluids}
  \bibinfo{volume}{53}~(\bibinfo{number}{1}) (\bibinfo{year}{2007})
  \bibinfo{pages}{57--80}.

\bibitem[{Gaskell and Lau(1988)}]{gask1988}
\bibinfo{author}{P.~H. Gaskell}, \bibinfo{author}{A.~K.~C. Lau},
  \bibinfo{title}{{Curvature-compensated convective transport: SMART, a new
  boundedness-preserving transport algorithm}}, \bibinfo{journal}{International
  Journal for Numerical Methods in Fluids}
  \bibinfo{volume}{8}~(\bibinfo{number}{6}) (\bibinfo{year}{1988})
  \bibinfo{pages}{617--641}.

\bibitem[{Patankar(1980)}]{patankar1980}
\bibinfo{author}{S.~V. Patankar}, \bibinfo{title}{{Numerical heat transfer and
  fluid flow}}, \bibinfo{publisher}{Taylor \& Francis}, \bibinfo{year}{1980}.

\bibitem[{Rhie and Chow(1983)}]{rhie1983}
\bibinfo{author}{C.~Rhie}, \bibinfo{author}{W.~Chow},
  \bibinfo{title}{{Numerical study of the turbulent flow past an airfoil with
  trailing edge separation}}, \bibinfo{journal}{AIAA Journal}
  \bibinfo{volume}{21}~(\bibinfo{number}{11}) (\bibinfo{year}{1983})
  \bibinfo{pages}{1525--1532}.

\bibitem[{Dongarra et~al.(1994)Dongarra, Lumsdaine, Pozo, and
  Remington}]{dongarra1994}
\bibinfo{author}{J.~Dongarra}, \bibinfo{author}{A.~Lumsdaine},
  \bibinfo{author}{R.~Pozo}, \bibinfo{author}{K.~Remington}, \bibinfo{title}{{A
  sparse matrix library in C++ for high performance architectures}}, in:
  \bibinfo{booktitle}{Proceedings of the Second Object Oriented Numerics
  Conference}, \bibinfo{pages}{214--218}, \bibinfo{year}{1994}.

\bibitem[{Botella and Peyret(1998)}]{botella1998}
\bibinfo{author}{O.~Botella}, \bibinfo{author}{R.~Peyret},
  \bibinfo{title}{Benchmark spectral solutions on the lid-driven cavity flow},
  \bibinfo{journal}{Computers \& Fluids}
  \bibinfo{volume}{27}~(\bibinfo{number}{4}) (\bibinfo{year}{1998})
  \bibinfo{pages}{421--33}.

\bibitem[{Ghia et~al.(1982)Ghia, Ghia, and Shin}]{ghia82}
\bibinfo{author}{U.~Ghia}, \bibinfo{author}{K.~N. Ghia}, \bibinfo{author}{C.~T.
  Shin}, \bibinfo{title}{{High-Re} solutions for incompressible flow using
  Navier-Stokes equations and a multigrid method}, \bibinfo{journal}{Journal of
  Computational Physics} \bibinfo{volume}{48} (\bibinfo{year}{1982})
  \bibinfo{pages}{387--411}.

\bibitem[{Fuchs and Tillmark(1985)}]{fuch1985}
\bibinfo{author}{L.~Fuchs}, \bibinfo{author}{N.~Tillmark},
  \bibinfo{title}{{Numerical and experimental study of driven flow in a polar
  cavity}}, \bibinfo{journal}{International Journal for Numerical Methods in
  Fluids} \bibinfo{volume}{5}~(\bibinfo{number}{4}) (\bibinfo{year}{1985})
  \bibinfo{pages}{311--329}.

\end{thebibliography}

\section*{Appendix A: The numerical method}
\begin{figure}[tb]
 \centering
 \begin{center}
 \includegraphics[width=0.5\textwidth]{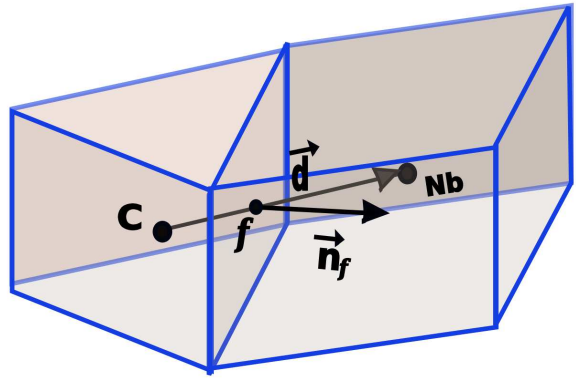}
\end{center}
 \caption{A computational cell $C$ and one of its neighbors $N_b$}
 \label{fig:cell}
\end{figure}

To describe the discretization practice used in the code, we can write the above equations (\ref{eq:mvt} and \ref{eq:energy}) in the generic convection-diffusion form:
\begin{equation}
 \dfrac{\partial}{\partial t} \int_{V}\rho \phi\ \dd V 
    + \int_{S} \rho \phi \vec U \cdot \vec n \ \dd S = 
    \int_{S} \Gamma \vec\nabla \phi\cdot \vec n \ \dd S + \int_{V} S_\phi\ \dd V
\label{eq:generic}\end{equation}
where $S_\phi$ is a source term. The spatial schemes approximating the diffusive and convective fluxes are both second-order accurate. The discretization of the diffusion term is performed by approximating the surface integrals by taking the sum over all the cell faces $f$ (see Figure \ref{fig:cell}):
\begin{equation}
 \int_{S} \Gamma \vec\nabla \phi\cdot \vec n \ \dd S = \sum_f \Gamma_f A_f (\vv{\nabla\phi})_f \cdot \vec n_f 
\end{equation}
In unstructured meshes, non-orthogonality is not an exception and it needs to be correctly handled. Thus, the normal gradient  $(\vv{\nabla\phi})_f \cdot \vec n_f$ is decomposed into an implicit contribution that uses the values of $\phi$ at the centers of the two cells sharing the face $f$ (the first term in the right hand side (RHS) of Eq. (\ref{eq:diffu}) ) and a non-orthogonality correction term treated explicitly by a deferred approach in order to preserve the second-order accuracy of this centered differencing. We use here the over-relaxed decomposition as suggested by \citet{jasak1996} to enhance the convergence properties of the diffusive term discretization:
\begin{equation}
 (\vv{\nabla\phi})_f \cdot \vec n_f = \dfrac{\phi_{N_b} - \phi_c}{||\vec d||}  \dfrac{1}{\vec d \cdot \vec n_f}+ \overline{\vv{ \nabla \phi}}
\cdot \left(\vec n_f - \frac{\vec d}{\vec d \cdot \vec n_f }\right)
\label{eq:diffu}
\end{equation}
$\vec d$ is the vector joining the centers of the two cells (see Figure \ref{fig:cell}). The average gradient $\overline{\vv{ \nabla \phi}}$ is interpolated from the gradients of these neighboring cells. 

The gradients of the variables at the cell centers are computed by Gauss' theorem:
\begin{equation}
\vv{ \nabla \phi} = \dfrac{1}{V} \sum_f \phi_f A_f\ \vec n_f
\end{equation}
where $\phi_f$ is the mean value of the variable using the values at the centers of two cells sharing the face $f$:
\begin{equation}
\phi_f = \xi \phi_c + (1-\xi) \phi_{N_b}\quad \text{with}\quad \xi = \dfrac{\overline{fN_b}}{\overline{CN_b} }
\end{equation}
 Once the gradient is calculated at all the computational cells, its new values are used to give a new estimate of $\phi_f$ as:
\begin{equation}
 \phi_f = \frac{1}{2} \left[ \left(\phi_{N_b} + \vv{ \nabla \phi}_{N_b} \cdot \vv{N_b f}\right) + \left(\phi_c + \vv{ \nabla \phi}_c \cdot \vv{C f} \right)\right]
\end{equation}
These new values of $\phi_f$ are used to recompute the gradients more accurately \citep{ferziger2002}.

Convection terms are also transformed to a sum over the faces $f$ composing the surface $S$:
\begin{equation}
\int_{S} \rho \phi \vec U \cdot \vec n \ \dd S = \sum_f (\rho\phi A)_f \vec U_f\cdot \vec n_f
\label{eq:8}
\end{equation}
where the face values $\phi_f$ require an appropriate interpolation to achieve accuracy and boundedness. For this purpose, we use the non-linear high resolution (HR) bounded scheme CUBISTA of \citet{alves2003}, in the $\gamma$ formulation of  \citet{ng2007}, where they express $\phi_f$ as a function of the upwind value of $\phi$  and of its centered differencing (CD) value:
\begin{equation}
 \phi_f^{HR} = \phi_{upwind} + \gamma(\phi_f^{CD}-\phi_{upwind})
\label{eq:HR}\end{equation}
The coefficient $\gamma$ is determined face by face based on the local shape of the flow solution, using the normalized variable diagram (NVD) framework and observing the convection boundedness criterion (CBC)\citep{gask1988}. The first term of the RHS of Eq. (\ref{eq:HR}) is accounted for implicitly, while the second term is treated explicitly with the deferred-correction practice.

The pressure-velocity coupling is ensured by the SIMPLE algorithm \citep{patankar1980}, while the mass fluxes at the cell faces are evaluated using the Rhie-Chow interpolation \citep{rhie1983} to avoid pressure checkerboarding. The implicit three-time step Gear's scheme of second-order accuracy is used to discretize the unsteady terms:
\begin{equation}
 \dfrac{\partial}{\partial t} \int_{V}\rho \phi\ \dd V 
= \dfrac{3(\rho \phi)_c^n - 4(\rho \phi)_c^{n-1} + (\rho \phi)_c^{n-2} }{\Delta t} V
\end{equation}
The superscript $n$ stands for the current time step and $\Delta t = t^n-t^{n-1}$ for the time step. The RHS of Eq. (\ref{eq:generic}) is taken at time $t^n$.

  At each iteration, the discretization technique presented above leads to a linear system with a non-symmetric sparse matrix for each variable. These linear systems are solved using an ILU-preconditioned GMRES solver using the implementation of the IML++ library \citep{dongarra1994}.

\section*{Appendix B: The code validation}

Different validation test cases have been performed to assess the accuracy of the code. We present here two cases corresponding to enclosed flows with sliding boundaries that make them similar to the flow in our two-rod mixer. They are both 2D flows. The first case is the classical lid-driven square cavity flow at a Reynolds number of $Re=1000$. Our findings are compared to the numerical results of \citet{botella1998}, which were obtained using a spectral Chebyshev method, and to those of \citet{ghia82}, who used a vorticity-stream function formulation. The second test case has a circular sliding boundary it is the polar lid-driven cavity flow studied experimentally and numerically by \citet{fuch1985}. We compare here our results to their experimental measurements obtained by laser Doppler anemometry at Reynolds numbers $Re=60$ and $350$.

\begin{figure}[tbh]
 \centering
 \begin{center}
 \includegraphics[width=0.41\textwidth]{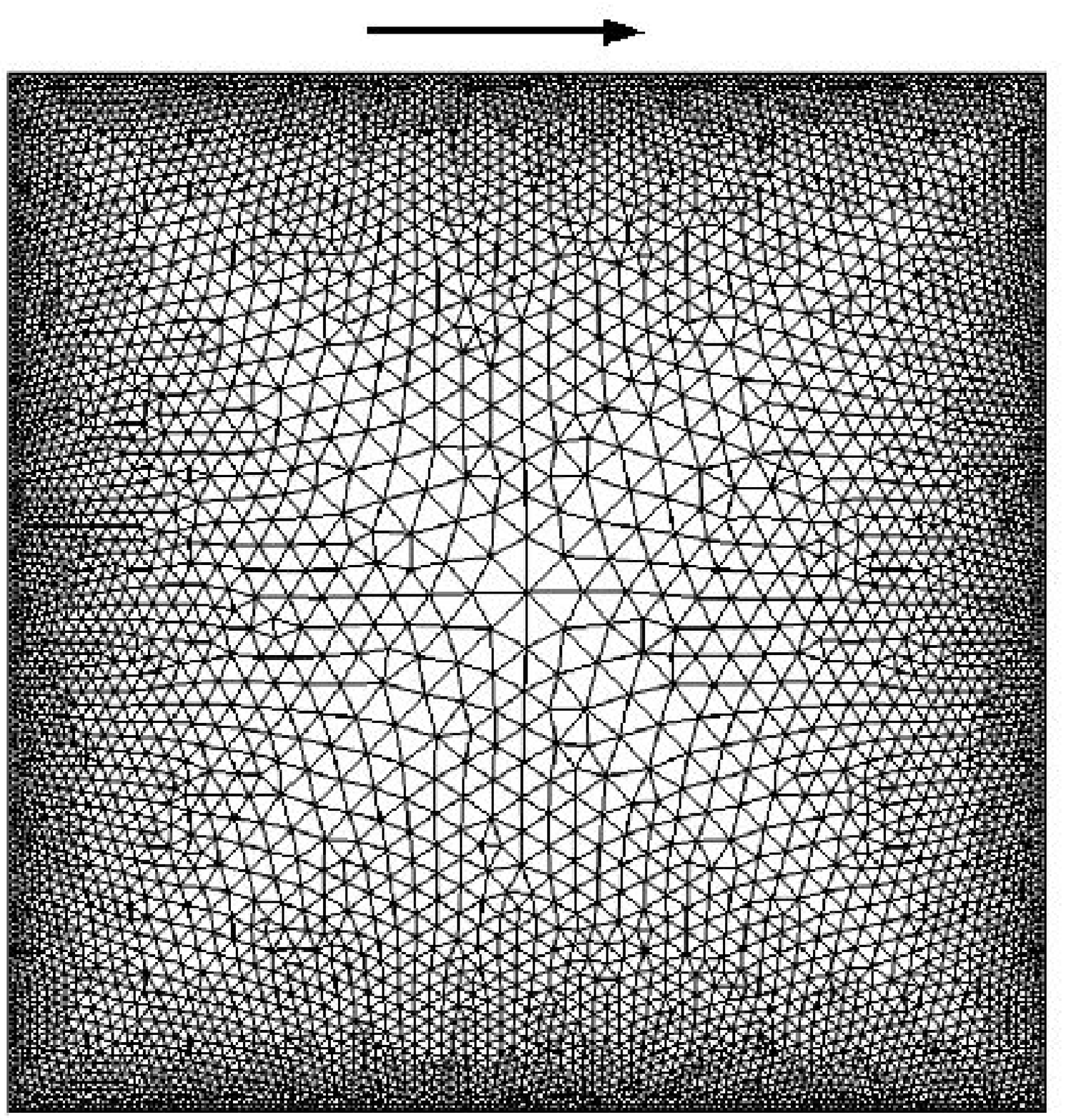}
 \includegraphics[width=0.58\textwidth]{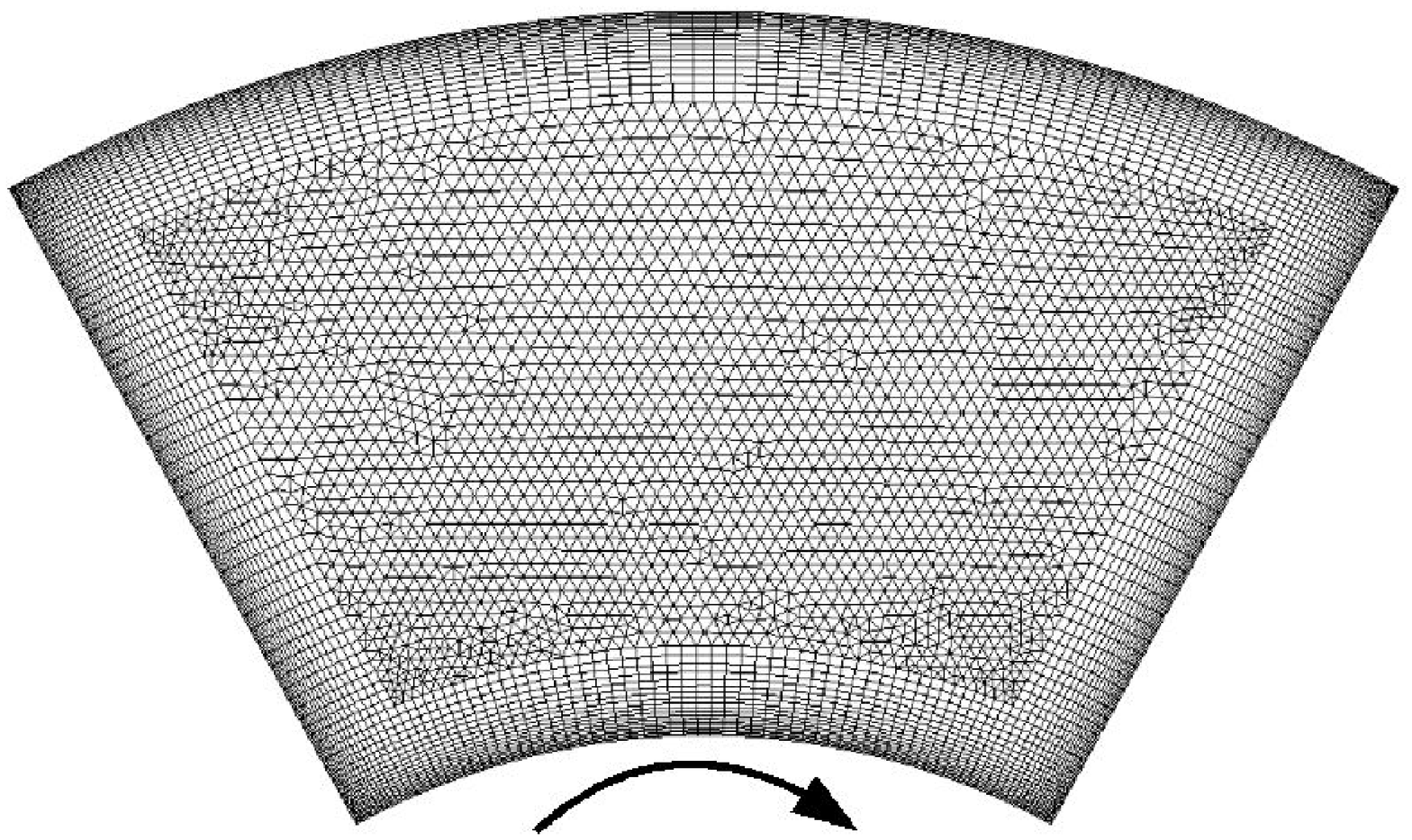}\\
(a)\hspace{5cm}(b)
\end{center}
 \caption{Computational meshes used for the validation. (a) A lid-driven square cavity and (b) a polar cavity with a rotating inner circular wall.}
 \label{fig:valid_mesh}
\end{figure}

The meshes used in the two cases are shown in Fig. \ref{fig:valid_mesh}. They are two-dimensional (\textit{i.e.} composed of one layer of cells) and of two different types. The mesh of the square cavity is composed of triangular-based prisms, only to demonstrate the ability of the code to handle non-orthogonal grids. It contains 10468 computational cells and is refined near the walls. The mesh of the polar cavity benefits from the hybrid mesh support and uses quadrilateral-based prisms (hexahedrals) near the boundaries to increase the mesh resolution and enhance the orthogonality, while triangular-based prisms are used elsewhere to ensure a flexible and homogeneous distribution of cells. This mesh is composed of 9610 cells.

\begin{figure}[tbh]
 \centering
 \begin{center}
 \includegraphics[width=0.49\textwidth]{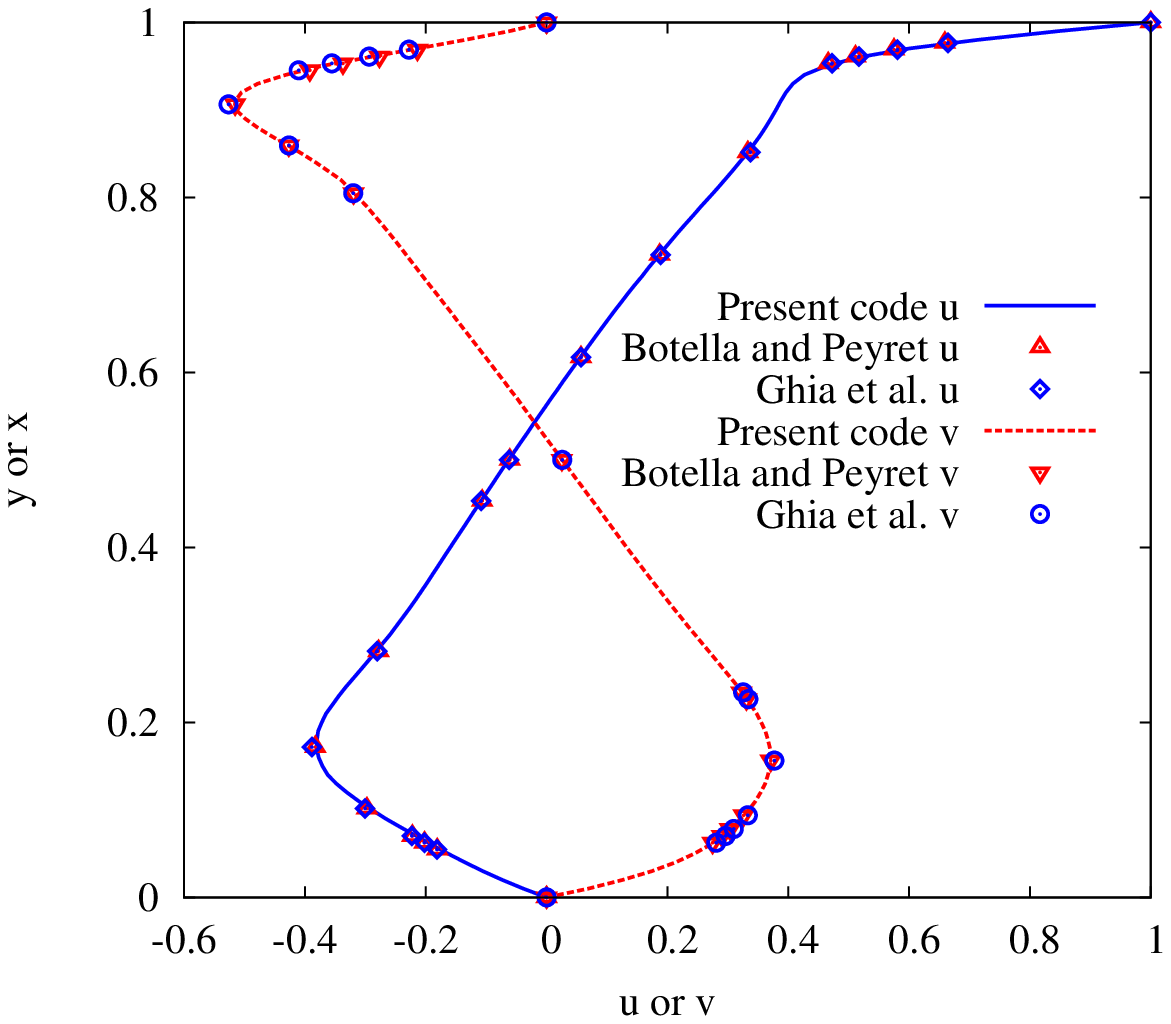}
 \includegraphics[width=0.49\textwidth]{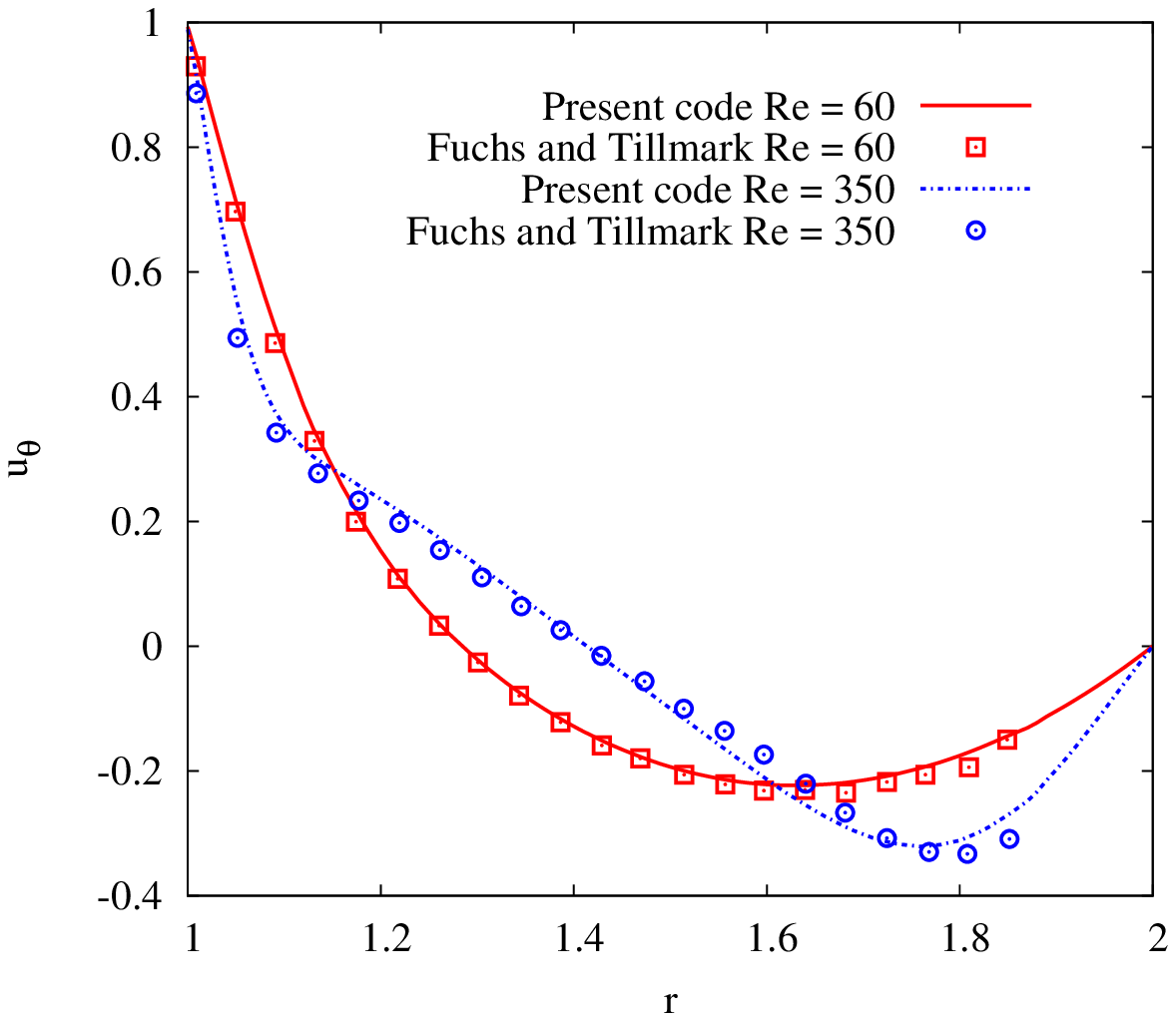}\\
(a)\hspace{5cm}(b)
\end{center}
 \caption{(a) The lid-driven square cavity: the $u$ and $v$ velocity components along the vertical and horizontal axes, $Re = 1000$. (b) The polar cavity: the $u_\theta$ velocity component along radial axis, $Re =60$ and $350$. }
 \label{fig:valid_results}
\end{figure}

In Fig. \ref{fig:valid_results} we present the obtained results and their comparisons. For the square cavity (Fig. \ref{fig:valid_results}(a)), the $x$ velocity component $u$ is plotted along the vertical central axis and the $y$ velocity component $v$ is plotted along the horizontal central axis. Our results match quite well with the benchmark results of \citet{botella1998} and \citet{ghia82}. Fig.~\ref{fig:valid_results}(b) shows the angular velocity along the central radial axis of the polar cavity. Good agreement is obtained with the experimental results of \citet{fuch1985}. Supplementary details and validation tests of this code are related in \citep{elomari2009a,elomari2010b}.

\end{document}